\documentclass{aa}

\usepackage{graphicx}
\usepackage{txfonts}
\usepackage{placeins} 
\usepackage{hyperref}

\begin{document}

   \title{A semi-analytical surrogate model for giant planet evolution}
    \subtitle{Bypassing ordinary
    differential equation solvers with localised thermodynamics, softplus asymptotes, and B-spline photometry}

   \author{C. Wilkinson\inst{1}\corrauth{christian.wilkinson@obspm.fr} 
          \and J.~Wehrung-Montpezat\inst{1}
          \and B.~Charnay\inst{1}
          \and M.~Mâlin\inst{2}
          \and L.~Delaye\inst{1}
          \and A.-M.~Lagrange\inst{1}
          \and V.~Squicciarini\inst{3}
          \and J.~Mazoyer\inst{1}
          \and S.~Mazevet\inst{4}
          \and B.~Perrier\inst{1}
          }

   \institute{
   LIRA, Observatoire de Paris, Universit\'{e} PSL, Sorbonne Universit\'{e}, Université Paris Cit\'{e}, CY Cergy Paris Universit\'{e}, CNRS, 92190 Meudon, France
    \and 
    Space Telescope Science Institute, 3700 San Martin Drive, Baltimore, MD 21218, USA
    \and
    Department of Physics \& Astronomy, University of Exeter, Stocker Road, Exeter, EX4 4QL, United Kingdom
    \and
    Observatoire de la Côte d'Azur, Universit\'{e} Côte d'Azur, 96, Boulevard de l'Observatoire, 06300 Nice, France
    }

   \date{Received 4 May 2026 / Accepted 10 August 2026}
    
    \abstract
    {Evolutionary models translate the observable luminosity, temperature, and colours of
    giant planets and brown dwarfs into mass and age. Generating their cooling tracks
    normally requires integrating the internal energy over time with an ordinary
    differential equation (ODE) solver coupled to pre-computed atmospheric grids, which
    becomes numerically stiff at sharp transitions such as cloud condensation and the
    onset of electron degeneracy, and is fragile inside Bayesian retrievals.}
    {We aim to generate continuous cooling tracks and photometric light curves directly from discrete atmospheric grids, without an ODE solver.}
    {We mapped the grids into a logarithmic thermodynamic space and extracted localised surrogate models with Gaussian-weighted regressions at fixed planetary parameters. We fitted the entropy and cooling rate against the internal temperature using bounded piecewise softplus functions to capture structural and cooling-rate transitions; the radius was fitted on the same temperature axis; and band photometry was represented with fixed-knot cubic B-splines. The age followed from numerical integration of these analytic functions, and uncertainties were propagated from the residual scatter of each fit.}
    {The surrogate, \texttt{CoolTrack}, reproduces the transition into electron degeneracy and the L-to-T spectral-type transition in the colour--magnitude diagram, converges on Solar System benchmarks, and evaluates a full evolutionary track in milliseconds on a standard desktop CPU.}
    {By removing the forward-modelling bottleneck, \texttt{CoolTrack} is suitable for direct use in Bayesian retrieval pipelines, where the age, mass, and formation entropy of a planet can be inferred jointly with its atmospheric properties.}

   \keywords{methods: analytical -- planets and satellites: physical evolution}

   \maketitle
   \nolinenumbers

\section{Introduction}

The discovery and characterisation of young giant exoplanets and brown dwarfs, primarily driven by direct imaging instruments and space missions (e.g. the Spectro-Polarimetric High-contrast Exoplanet REsearch instrument at the Very Large Telescope, VLT/SPHERE; the Gemini Planet Imager, Gemini/GPI; and JWST), rely heavily on theoretical evolutionary models \citep{burrows_nongray_1997, chabrier_theory_2000, spiegel_deuterium-burning_2011}. Because these substellar objects do not fuse hydrogen, they lack a stable internal energy source. Consequently, their observable properties, such as bolometric luminosity, effective temperature, and photometric emission, are entirely dictated by the progressive loss of their primordial heat over time. By comparing a measured planetary flux to a theoretical cooling track, observers can infer fundamental planetary parameters, bridging the critical gap between an object's measured age and its true mass. This method has historically underpinned the characterisation of benchmark systems such as HR~8799 \citep{marois_direct_2008} and $\beta$~Pictoris~b \citep{lagrange_probable_2009}. The same cooling-track framework applies to the isolated substellar objects below the deuterium-burning mass limit identified in young clusters and the field \citep{osorio_discovery_2000, kirkpatrick_new_2005, luhman_formation_2012, caballero_review_2018}. As observational capabilities have advanced, evolutionary models continue to play a crucial role, recently enabling the identification of the sub-Jovian candidate TWA~7~b with JWST \citep{lagrange_evidence_2025} and the mass inference of HD~143811~(AB)b using Gemini/GPI \citep{squicciarini_gpisphere_2025}.

For over two decades, comparing observed luminosities to theoretical cooling tracks has been the standard paradigm for inferring the masses of directly imaged substellar companions \citep{baraffe_evolutionary_2003, chabrier_giant_2014, bowler_imaging_2016}. Physically, the evolution of these fully convective objects is governed by a monotonic decrease in internal specific entropy ($S$). As the planet radiates heat into space, it cools and undergoes gradual structural contraction \citep{burrows_nongray_1997}. A young, inflated planet initially contracts following ideal gas laws until its deep interior reaches densities high enough to trigger electron degeneracy. This quantum mechanical pressure halts further macroscopic contraction, stabilising the planet at a relatively constant radius for the remainder of its lifetime.

Traditionally, generating one of these evolutionary cooling tracks has required coupling a pre-computed atmospheric boundary condition grid, such as the ATMO \citep{phillips_new_2020}, Sonora \citep{marley_sonora_2021}, or \texttt{HADES} \citep{wilkinson_breaking_2024} models, to an internal structural model to dictate how efficiently heat escapes the atmosphere. The planet's age is then derived by numerically integrating the energy equation over time using an ordinary differential equation (ODE) solver. 

However, modern atmospheric grids are complex. They incorporate multidimensional parameters such as varying metallicities ($Z$), stellar irradiation (the radiative-equilibrium temperature, $T_{\rm irr}$, is set by the absorbed stellar flux; \citealt{wilkinson_breaking_2024}), and intricate cloud microphysics \citep{baudino_interpreting_2015, charnay_self-consistent_2018}. Physical phenomena, such as the sudden condensation of a thick cloud deck, introduce sharp gradients in the tabulated cooling derivatives ($\mathrm{d}S/\mathrm{d}t$), rendering the underlying energy equations stiff.

Comprehensive benchmarking of numerical integration methods reveals that while specialised algorithms (e.g. backward differentiation formulas; \citealt{curtiss_integration_1952}) can navigate stiff forward simulations, they are sensitive to error tolerances and linear solver configurations \citep{stadter_benchmarking_2021}. Crucially, this numerical stiffness is severely exacerbated during Bayesian parameter inference. As a Markov chain Monte Carlo (MCMC) sampler explores the posterior landscape, it inevitably evaluates non-physical parameter combinations that trigger pathological stiffness far beyond what is encountered in isolated forward modelling \citep{stadter_benchmarking_2021}.

Compounding this computational fragility, traditional ODE solvers introduce subtle numerical artefacts when deployed within these inverse frameworks. As demonstrated by recent numerical studies \citep{creswell_understanding_2023}, ODE solvers can imperceptibly corrupt the likelihood surfaces required for Bayesian retrieval. For fixed-step solvers, coarse integration steps introduce systematic truncation errors that physically shift the likelihood surface, biasing the converged parameter posteriors. More severely, adaptive-step solvers automatically adjust their step sizes and integration sequences for slightly different parameter evaluations. These discrete algorithmic jumps generate artificial, spurious discontinuities in the likelihood landscape, creating solver-induced phantom local optima that trap MCMC samplers.

This dual threat of mechanical stiffness and solver-driven likelihood distortion creates a severe computational bottleneck. It effectively precludes the use of dynamic, on-the-fly evolutionary tracking within modern retrieval pipelines \citep{madhusudhan_temperature_2009, molliere_petitradtrans_2019, al-refaie_taurex_2021, macdonald_poseidon_2023, wilkinson_breaking_2024}. This barrier has recently driven the field towards machine-learning surrogate models and neural-network emulators, which require large pre-computed training sets \citep{himes_accurate_2022, ardevol_martinez_floppity_2024}, and polynomial surrogate frameworks that achieve similar efficiency with far fewer evaluations \citep{de_wringer_surrogate-accelerated_2026}.

In this work, we propose a semi-analytical alternative. We present \texttt{CoolTrack}, an evolutionary engine designed to bypass the traditional ODE solver bottleneck entirely. Rather than integrating step by step through a discrete numerical grid, it extracts continuous, localised surrogate models of the planet's fundamental physics using distance-weighted regressions in a logarithmic phase space. By replacing complex grid interpolations with smooth, analytical equations, specifically utilising bounded piecewise softplus functions (curves that transition smoothly between linear and flat asymptotic regimes) for structural contraction and cooling rate transitions, and cubic B-splines (flexible, localised polynomial curves) for non-linear photometric transitions, it guarantees mathematical stability and achieves computational efficiency across broad planetary parameter spaces.

\section{Methodology}

\subsection{Structural grid comparison}
\label{Sec: grid}

Before extracting the semi-analytical surrogate models, it is essential to contextualise our raw numerical base grid, \texttt{HADES}, against established literature models. Extensive clear-sky validations of the \texttt{HADES} structural solver against ATMO \citep{phillips_new_2020}, and Sonora \citep{marley_sonora_2021} baselines were previously demonstrated by \citet{wilkinson_breaking_2024}. Therefore, in this work, we focused exclusively on modelling realistic, cloudy atmospheres. 

The evolutionary \texttt{CoolTrack} engine relies on a production grid incorporating a multi-species cloud microphysics model with $100\%$ spatial coverage. As a giant planet cools over billions of years, we explicitly modelled five distinct condensate clouds to capture true opacity transitions across the cooling sequence \citep{mason_chemistry_2006, visscher_atmospheric_2010}. The internal (intrinsic) temperature ($T_{\rm int}$) is defined through the intrinsic flux ($\sigma T_{\rm int}^4$) radiated from the interior; it differs from the effective temperature, which also carries reprocessed irradiation, via $T_{\rm eff}^4 = T_{\rm int}^4 + T_{\rm irr}^4$ \citep{wilkinson_breaking_2024}. Iron (Fe) and forsterite (Mg$_2$SiO$_4$) govern the high-temperature regime ($T_{\rm int} \gtrsim 1500$~K) where refractory silicate and iron species condense from the gas phase \citep{lodders_alkali_1999, visscher_atmospheric_2010}. Water (H$_2$O), ammonium hydrosulfide (NH$_4$SH), and ammonia (NH$_3$) clouds sequentially precipitate in the mature, deeply cooled regime ($T_{\rm int} \lesssim 400$~K), mirroring the condensation sequence observed in the outer Solar System giant planets \citep{burrows_nongray_1997, morley_thermal_2015}; ammonia clouds in particular leave a visible imprint on the cooling tracks \citep{chen_jupiter_2023}. Intermediate salt and sulfide clouds (e.g. KCl and Na$_2$S) were omitted from the baseline grid, as our ablation tests showed that they introduced severe numerical instabilities during atmospheric convergence without significantly altering the bulk cooling rate.

To maintain physical realism across these distinct condensation regimes while avoiding the curse of dimensionality associated with fully independent cloud parameters, we split the sedimentation efficiency ($f_{\rm sed}$) into two distinct, freely varying parameters. The parameter $f_{\rm sed,ref}$ exclusively governs the deep, high-temperature clouds (Fe and Mg$_2$SiO$_4$), while $f_{\rm sed,vol}$ controls the upper-atmosphere, low-temperature clouds (H$_2$O, NH$_3$, and NH$_4$SH).

This comprehensive microphysics introduces a strong atmospheric blanketing (greenhouse) effect. The thick cloud decks trap internal heat, severely suppressing the cooling rate and shifting the deep radiative-convective boundary to higher pressures and temperatures. Consequently, the fully cloudy planetary structure maintains a systematically larger, more inflated physical radius at a given internal temperature ($T_{\rm int}$) compared to clear-sky environments. 

\begin{figure*}[ht]
    \centering
    \includegraphics[width=\textwidth]{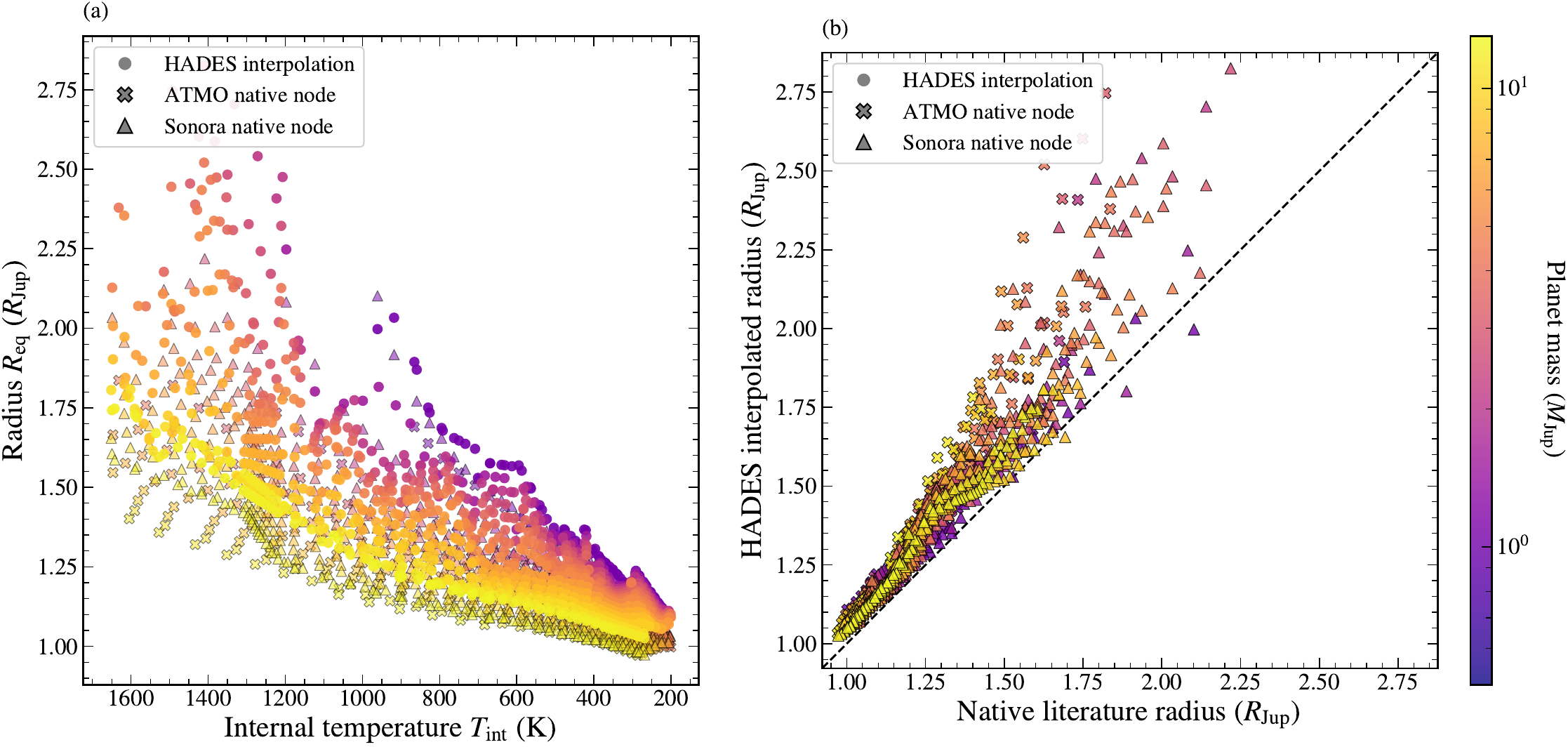}
    \caption{State-space structural mapping comparing the fully cloudy \texttt{HADES} grid ($100\%$ coverage, five cloud species) evaluated at the native $(M, T_{\rm int})$ nodes of the clear-sky ATMO and Sonora literature models. The mass domain is bounded between $0.5$ and $13.0~M_{\rm Jup}$. The strong atmospheric greenhouse effect from the full cloud decks naturally inflates the \texttt{HADES} radii, shifting the points consistently above the 1:1 dashed parity line, in contrast to the clear-sky baselines.}
    \label{fig:grid_comparison}
\end{figure*}

\begin{figure*}[ht]
    \centering
    \includegraphics[width=\textwidth]{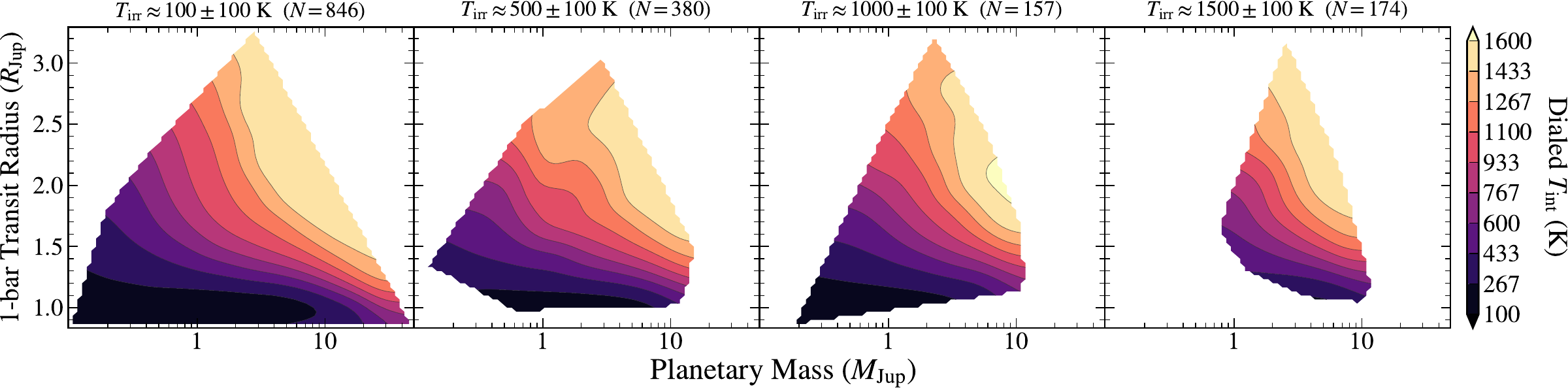}
    \caption{Smoothed topological contours of the \texttt{HADES} grid mass--radius relationship, evaluated across four distinct stellar irradiation bins ($T_{\rm irr} \approx 100, 500, 1000,$ and $1500$~K). The colour map represents the internal temperature ($T_{\rm int}$). The strict grid filtering ensures a fixed parameter space (solar metallicity, $10~M_\oplus$ core mass, and baseline cloud sedimentation), cleanly highlighting the physical inflation driven by increasing internal heat and external stellar irradiation (contours are lightly Gaussian-smoothed for clarity).}
    \label{fig:mass_radius_contours}
\end{figure*}

This physical inflation is illustrated in Fig.~\ref{fig:grid_comparison}, which compares the fully cloudy \texttt{HADES} grid ($100\%$ coverage, five cloud species) evaluated exactly at the native mass and temperature nodes of the clear-sky literature baselines. The expected radial divergence driven by the greenhouse effect is visible, shifting the cloudy radii significantly above the 1:1 parity line. Furthermore, the broad structural topology of our filtered base grid is visualised in Fig.~\ref{fig:mass_radius_contours}, showcasing the smooth mass--radius relationships parameterised by internal heat across changing irradiation environments.

The core mechanism of the \texttt{CoolTrack} framework relies on transforming the discrete, multidimensional atmospheric grids evaluated in Sect.~\ref{Sec: grid} into continuous, differentiable surfaces. Rather than attempting to integrate across these tabulated boundaries directly, we projected the raw grid data into a fully logarithmic thermodynamic phase space. Within this stabilised domain, we identified analytical relationships between internal temperature ($T_{\rm int}$), specific physical entropy ($S$), inverse cooling rate or residence time ($\tau$), planetary radius ($R_{\text{eq}}$), and photometric flux ($F_{\lambda}$). Here $R_{\text{eq}}$ is the radius at the 1-bar level of the spherically symmetric \texttt{HADES} structure; rotation and the associated oblateness (the Theory of Figures) are not modelled, so it is a mean radius rather than a true equatorial radius. The structures assume a hydrogen--helium envelope (protosolar helium fraction) at solar metallicity above a $10\,M_\oplus$ core, computed with the \citet{chabrier_new_2021} equation of state.

\subsection{Localised Gaussian weighting}

A key design choice of the \texttt{CoolTrack} framework is the use of randomised sampling rather than the regular, evenly spaced grids traditionally employed in pre-computed atmospheric model libraries \citep{charnay_self-consistent_2018, phillips_new_2020, marley_sonora_2021}. Regular grids implicitly assume that physical quantities vary smoothly and uniformly across parameter space, making linear interpolation between grid points a reasonable approximation. However, when the underlying physics contains sharp transitions of unknown position and width, such as the abrupt condensation of cloud decks or the onset of electron degeneracy, the exact location and spacing of these features cannot be known a priori. In such cases, a regular grid risks systematically misrepresenting or entirely missing physical transitions that fall between rigid grid coordinates. 

By drawing grid points across the parameter bounds, the \texttt{CoolTrack} engine makes no prior assumption about where physical transitions occur. To ensure an unbiased and gap-free sampling strategy, we utilised Latin hypercube sampling \citep{mckay_comparison_1979}. This scheme is a stratified randomised approach that forces the sampling algorithm to test every sub-interval of a parameter's range exactly once. As demonstrated by \citet{mckay_comparison_1979}, the sampling guarantees the full marginal coverage of a structured grid without the exponential $N^K$ computational cost (where $N$ is the number of samples per parameter and $K$ is the number of independent dimensions), yielding tighter variance bounds than simple Monte Carlo random sampling and preventing localised parameter clustering. To properly resolve the non-linear evolution of the low-mass regime, we computed this sampling for the planetary-mass axis in logarithmic space.

To extract the localised surrogate models for a target planet defined by a specific parameter vector (e.g. $M$, $Z$, and $T_{\rm irr}$), we computed a localised influence weight for every point in the pre-computed grid using a Gaussian smoothing kernel \citep[e.g.][]{cleveland_locally_1988}. Because the physical scales of the independent variables differ by orders of magnitude, all independent dimensions are first standardised to zero mean and unit variance (i.e. expressed as z-scores). For the mass axis, this standardisation is performed in $\log$ space. 

The relative weight ($w_i$) for each grid point is then assigned based on its Euclidean distance ($d_i$) from the target in this standardised multidimensional space:
\begin{equation}
    w_i \propto \exp\left( -\frac{d_i^2}{2\sigma^2} \right)
,\end{equation}
where $\sigma$ acts as the bandwidth parameter, defaulted to $0.5$ standard deviations. These proximity weights are subsequently normalised such that $\sum w_i = 1$ prior to regression.

To ensure the regressions capture sufficient local physics while preventing over-smoothing from distant, irrelevant atmospheric states, the effective number of models per regression is dynamically bounded. First, the Gaussian tail is strictly truncated by rejecting any grid points with $w_i < 0.05$. Because a Gaussian distribution never naturally decays to zero, failing to truncate the tail would force the non-linear solver to evaluate thousands of distant grid points that contribute negligible statistical weight. A threshold of $0.05$ corresponds mathematically to a Euclidean distance of approximately $2.45\sigma$ from the target. Enforcing this finite-support boundary reduces the computational overhead of the solver and explicitly shields the local regression from the influence of physically distinct atmospheric regimes (such as crossing a distant cloud-condensation threshold). However, to ensure the resulting design matrices remain fully determined for our multi-parameter non-linear fits, an algorithmic safety net is enforced: if local grid sparsity results in fewer than five valid grid points within the truncation threshold, the engine defaults to utilising the 50 nearest neighbours regardless of weight. This dual-bound approach guarantees stable matrix inversion while aggressively limiting non-local diversity.

\subsection{The internal-temperature coordinate}

All surrogate relations in \texttt{CoolTrack} are parameterised by a single evolutionary coordinate, the internal temperature ($T_{\rm int}$). The entropy, $S(T_{\rm int})$, the inverse cooling rate, $\tau(T_{\rm int})$, the structural radius, $R_{\text{eq}}(T_{\rm int}),$ and the band photometry, $F_\lambda(T_{\rm int}),$ are each extracted as functions of $\ln T_{\rm int}$. The entropy then enters only through the age integral (Sect.~\ref{Sec:fomula}), where $t = -\int \tau\,\mathrm{d}S$. Photometric flux is naturally a function of $T_{\rm int}$ (and of the background field $T_{\rm irr}$), since it is the surface emission escaping the photosphere.

A natural objection is that $S$ and $T_{\rm int}$ are not globally one-to-one: at fixed entropy, two planets of different surface gravity or cloud thickness radiate at different internal temperatures, so the radius, a bulk property set by the equation of state and the onset of electron degeneracy, is most directly a function of $S$. This holds across the grid as a whole, but not within the localised neighbourhood used for a single surrogate extraction. Because the Gaussian proximity weights collapse the fit onto a narrow region of the static parameter vector $\boldsymbol{\theta}=(M, Z, T_{\rm irr}, M_{\text{core}}, f_{\text{sed,ref}}, f_{\text{sed,vol}})$, and in particular onto a narrow range of mass, and hence gravity, the mapping between $S$, $T_{\rm int}$, and $R_{\text{eq}}$ is effectively single-valued over the points that carry statistical weight. Fitting $R_{\text{eq}}(T_{\rm int})$ locally is therefore well posed, and it reproduces the mass-clean $R(T_{\rm eff})$ relation that the underlying \texttt{HADES} grid was validated against, keeping $R(\mathrm{age})$ consistent with $T_{\rm eff}(\mathrm{age})$.

The decisive advantage is in the uncertainty budget. Were the radius fitted against $S$, with $S$ itself a fitted surrogate, each radius sample would inherit two stochastic layers, the entropy draw and the radius draw, and the Monte Carlo envelope would compound them, inflating the radius band well beyond the quality of the underlying fit. Parameterising the radius on the same $T_{\rm int}$ axis as every other quantity removes this double mapping: the radius uncertainty (Sect.~\ref{Sec:fomula}) is evaluated independently of the entropy and cooling-rate draws, so the age uncertainty and the radius uncertainty stay orthogonal, yielding clean confidence intervals in the radius--age plane.

\subsection{Localised analytical surrogate extraction}
\label{Sec:fomula}
\begin{figure*}[ht]
    \centering
    \includegraphics[width=\textwidth]{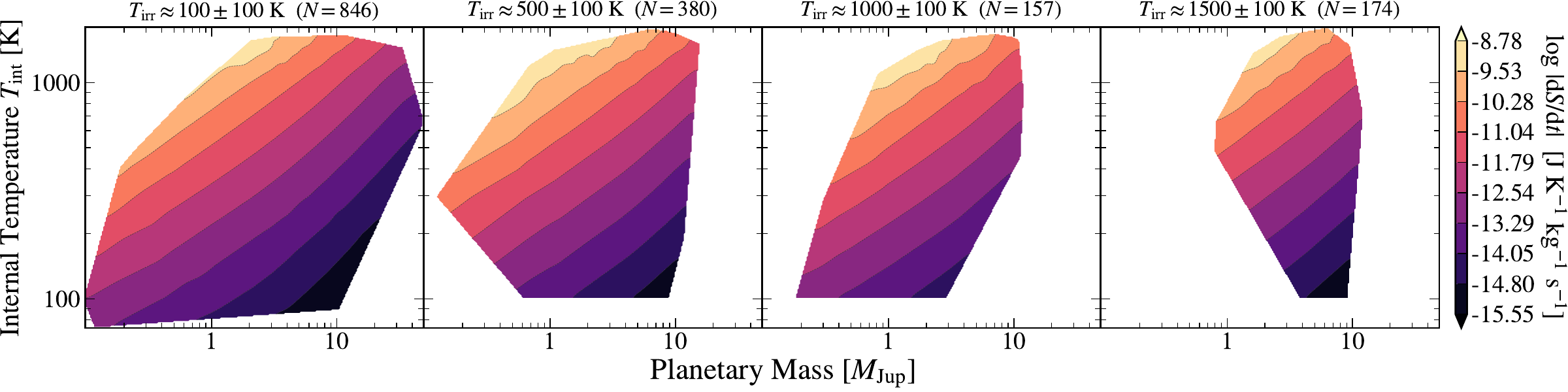}
    \caption{Topological contours of the planetary cooling rate magnitude ($\log|\mathrm{d}S/\mathrm{d}t|$) across four distinct stellar irradiation regimes. The data are projected across the parameter space of planetary mass and internal temperature ($T_{\rm int}$). The smooth, parallel contour lines visually demonstrate that the cooling rate is overwhelmingly governed by mass and internal heat. While high irradiation ($T_{\rm irr} = 1500$~K) visibly shifts the contours, representing the suppression of heat loss via the atmospheric greenhouse effect, the underlying log-linear dependences remain stable, validating the use of localised log-linear surrogate regressions within the \texttt{CoolTrack} engine. Contours are lightly Gaussian-smoothed for clarity.}
    \label{fig:cooling_rate_dependencies}
\end{figure*}

Following the enforcement of top-of-atmosphere energy balance to correctly extract internal heat free from deep-atmospheric convective bottlenecks, we find that the fundamental thermodynamics behave log-linearly across the vast majority of the evolutionary sequence. This log-linear behaviour is empirically justified by the global phase space; as demonstrated in Fig.~\ref{fig:cooling_rate_dependencies}, the general topology is overwhelmingly dominated by internal temperature and planetary mass. However, during major phase transitions, the relationship between surface emission and deep structural entropy can diverge. Because the Gaussian proximity weights are recalculated for every target evaluation, the resulting regression coefficients are explicitly dependent on the target planet's static parameter vector, $\boldsymbol{\theta} = (M, Z, T_{\rm irr}, M_{\text{core}}, f_{\text{sed,ref}}, f_{\text{sed, vol}})$. To account for thermodynamic lags caused by abrupt cloud clearing, we implemented an adaptive framework for the fundamental thermodynamics ($\ln S$ vs $\ln T_{\rm int}$). The algorithm dynamically tests a bounded piecewise softplus function to capture structural knees, sharp inflections in the cooling slope, safely reverting to a locally weighted linear regression if no significant physical transition is present:

\begin{equation}
    \ln S = \alpha_S(\boldsymbol{\theta}) \ln(T_{\rm int}) + \beta_S(\boldsymbol{\theta})
    \label{eq:loglinear}
.\end{equation}
Here $\alpha_S$ and $\beta_S$ are the localised regression coefficients when log-linear behaviour dominates. For tracks extending into the deeply cooled regime ($T_{\rm int} \lesssim 300$~K), where a second inflection appears as the low-temperature volatiles condense, the algorithm promotes the entropy fit to a two-knee (dual-softplus) form; otherwise, it retains the single-knee softplus or the log-linear regression of Eq.~(\ref{eq:loglinear}).

Similarly, the inverse cooling rate ($\ln \tau$, defined as the planet's thermal residence time in seconds per specific entropy unit, i.e., $\tau = |\text{d}t/\text{d}S|$) exhibits a distinct bifurcation. As a cloudy planet cools through the L-to-T transition ($T_{\rm int} \approx 900$--$1400$~K), the shedding of the thick volatile cloud decks forces the object to radiate away a large amount of trapped internal heat. This creates a distinct plateau in the cooling sequence \citep{saumon_evolution_2008}, separating a steep high-temperature cloudy cooling slope from a shallower low-temperature clear-sky cooling slope.

A simple linear regression across this domain mathematically erases this transition. Therefore, we dynamically modelled the inverse cooling rate ($\ln \tau$) using a bounded piecewise softplus function, anchoring the transition knee ($x_0$) to the physical cloud-clearing temperature window. This anchoring lets the surrogate bridge the cloudy and cloudless cooling regimes. Crucially, the mathematical derivative of this softplus transition can naturally reproduce the statistical pile-up (residence time spike) of transitional brown dwarfs predicted by numerical models \citep[e.g.][]{saumon_evolution_2008}, ensuring the surrogate captures the self-consistent boundary physics of cloud clearing. If the local grid data reflect a purely clear-sky or single-regime evolution lacking this plateau, the algorithm dynamically reverts to a standard log-linear fit for $\ln \tau$.

Deep structural contraction similarly transitions sharply but continuously when electron degeneracy pressure overtakes ideal gas laws. Simple linear or power-law approximations are poorly suited to capturing this distinct knee. To ensure mathematical stability across this phase transition, we modelled the structural radius ($R_{\text{eq}}$) using a secondary bounded piecewise softplus function:
\begin{equation}
    \ln R_{\text{eq}} = y_0 + k_1(x - x_0) + \frac{k_2 - k_1}{\beta} \ln\left(1 + \exp(\beta(x - x_0))\right)
,\end{equation}
where $x$ represents the internal temperature $\ln T_{\rm int}$. Rather than relying on discrete piecewise models, this formalism provides a smooth, analytically differentiable transition between two distinct physical asymptotes, establishing the explicit modelling of electron degeneracy as a native possibility within the framework. The parameter $x_0$ isolates the critical transition point into degeneracy, $k_1$ and $k_2$ represent the distinct structural scaling slopes of the ideal gas and degenerate regimes, and $\beta$ acts as a fitted sharpness parameter to naturally adapt to the transition rate. The validity of this analytical choice is empirically demonstrated by the acceptable fits to the underlying discrete grid data ($R^2 \ge 0.86$), ensuring the cooling derivatives remain continuous as the planet's core becomes degenerate.

While planetary structural contraction behaves asymptotically at the degeneracy limit, photometric emission ($F_{\lambda}$) frequently exhibits complex, non-linear phase transitions caused by the sequential condensation and clearing of atmospheric cloud decks, as well as the shifting peak of the Planck function \citep[e.g.][]{burgasser_spectra_2002, ackerman_precipitating_2001}. To map these sharp observable light curves while enabling analytical error propagation, we modelled the photometric fluxes using fixed-knot cubic B-splines as a function of internal temperature, $\ln T_{\rm int}$. Conceptually, a B-spline is a sequence of localised, simple polynomial curves joined together at specific points called knots.

We explicitly chose this localised approach over fitting a single, global polynomial across the entire temperature range. Global polynomials are susceptible to Runge's phenomenon, a numerical instability where forcing the function to fit a sharp, local feature, such as the high-temperature $J$-band hook, a rapid photometric reversal driven by cloud evolution, causes unphysical oscillations at the boundaries of the evolutionary curve. B-splines natively combat this by confining the mathematical influence of any sharp transition strictly to its local knots.

Because a cubic B-spline is constructed from a linear combination of strictly localised polynomial basis functions, it provides infinite flexibility with local control. The mathematical formulation takes the form
\begin{equation}
    \log F_{\lambda}(x) = \sum_{i=0}^{n} c_i B_{i,3}(x)
,\end{equation}
where $x = \ln T_{\rm int}$, $n+1$ is the total number of control points, $c_i$ are the fitted control point coefficients, and $B_{i,3}(x)$ are the cubic basis functions defined over a fixed knot vector. In our implementation, we set $n=5$, resulting in exactly six control points. This localised formulation allows the surrogate to adapt to flat background irradiation floors, sharply climb through cloud-clearing transitions, and capture high-temperature inversions without disrupting the rest of the evolutionary curve.

\begin{figure*}[ht]
    \centering
    \includegraphics[width=\textwidth]{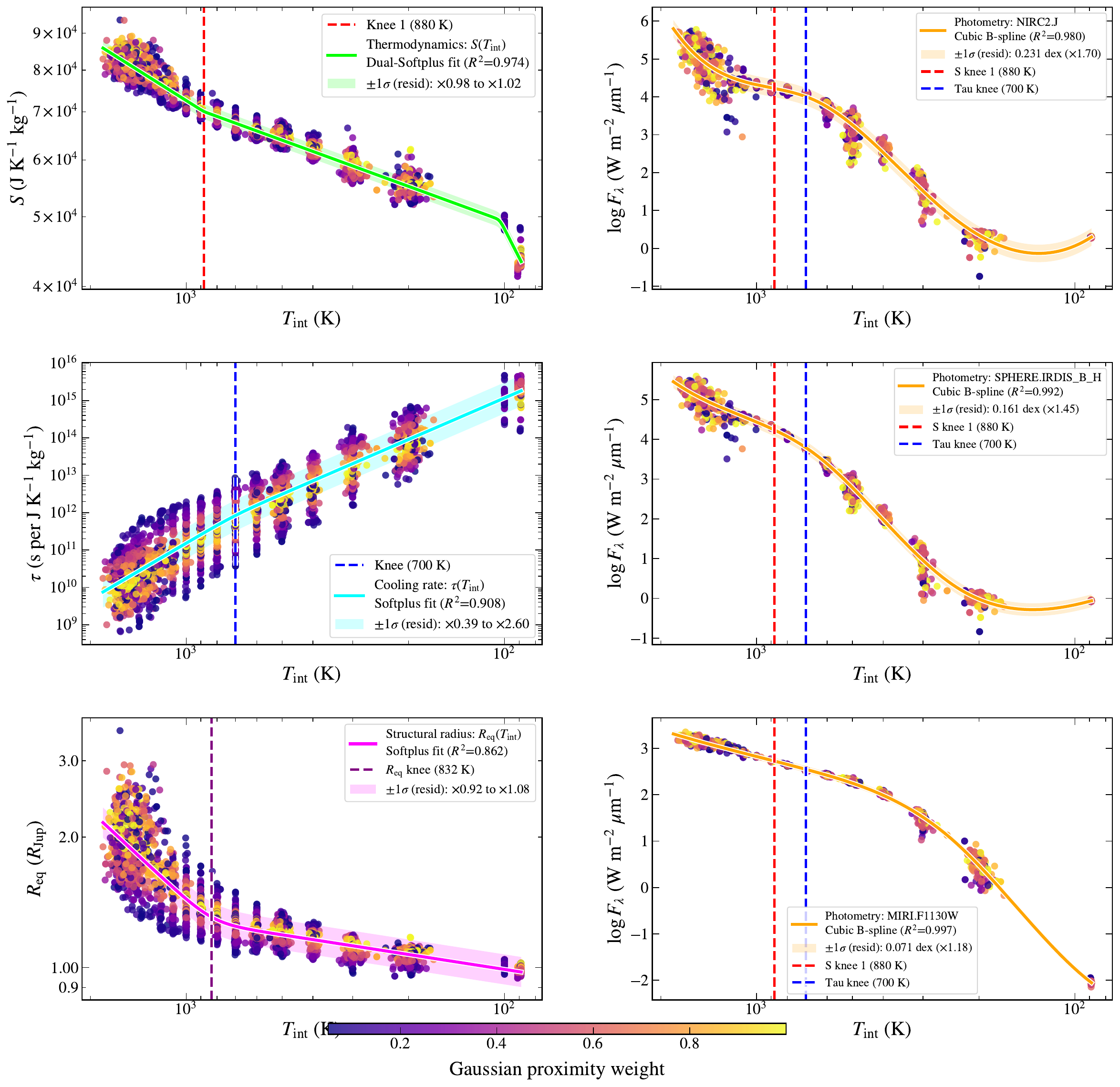}
    \caption{Comprehensive analytical surrogate model extraction evaluated for a representative target planet. The discrete \texttt{HADES} grid points are plotted such that their colour and opacity correspond directly to their multidimensional Gaussian proximity weight, allowing the regressions to dynamically isolate local physics. \textit{Left column}: Fundamental physical surrogates (from top to bottom): (i) Thermodynamics ($\ln T_{\rm int}$ vs $\ln S$) and (ii) cooling rate ($\ln T_{\rm int}$ vs $\ln \tau$), each fitted with an adaptive softplus (or, where the track extends below $\sim 300$~K, a two-knee dual softplus) and reverting to a log-linear fit where no transition is present; and (iii) the structural equation of state ($\ln T_{\rm int}$ vs $\ln R_{\text{eq}}$), modelled with a bounded piecewise softplus to capture the onset of electron degeneracy. \textit{Right column}: Photometric emission surrogates in three representative photometric bands (Keck/NIRC2 $J$, VLT/SPHERE $H$, and JWST/MIRI F1130W), from top to bottom. The fluxes are modelled with cubic B-splines to capture the multi-regime cloud-clearing transitions. Shaded bands on every panel are the proximity-weighted residual scatter of each fit. The surrogate framework adapts to the localised physics, yielding high weighted coefficients of determination ($R^2$) across all domains. The vertical dashed lines mark the extracted physical knees, mathematically identifying critical transition thresholds, such as the onset of electron degeneracy and the shedding of volatile clouds, isolated directly from the softplus asymptotes.}
\label{fig:surrogate_extraction}
\end{figure*}

Crucially, this approach differs from how we treated the structural and thermodynamic regimes. We relied on the bounded piecewise softplus function for the structural equation of state because it mathematically enforces the known physical asymptotes of ideal gas contraction and electron degeneracy. B-splines are purely empirical and lack these physical bounds, making them ideal for the complex morphology of photometry but inappropriate for enforcing fundamental structural limits. Furthermore, because the B-spline optimisation is linear in its coefficients, the fit is stable across sparse boundaries and no longer requires the nearest-neighbour interpolation previously used to fill flux gaps; a $k$-d tree is retained only for the localised coefficient lookup in the pre-compiled predictor, not for interpolating fluxes. The complete, scalable surrogate extraction process, spanning thermodynamics, cooling rate, structural radius, and photometric fluxes, is illustrated in Fig.~\ref{fig:surrogate_extraction}. The unified 6-panel dashboard demonstrates the framework's ability to smoothly map isolated physical regimes across all physical dimensions.

\subsection{Integration and initial conditions}
With the thermodynamics and structural evolution defined entirely by continuous and localised analytical functions, the exact age of the planet can be derived via rapid numerical integration. To begin this integration, the engine requires a starting specific physical entropy, $S_0$. 

We determined this initial boundary condition dynamically by utilising the theoretical post-formation mass--entropy distributions derived by \citet{mordasini_characterization_2017}. Specifically, we extracted the upper and lower entropy bounds as a function of planetary mass, which correspond to the classical hot-start (high-entropy) and cold-start (low-entropy) formation scenarios \citep{marley_luminosity_2007} as well as intermediate bounds, the classical warm-start \citep{mordasini_characterization_2017}. Our framework constructs continuous interpolators across these mass-dependent boundaries, extracting the appropriate starting entropy, $S_0$.

Because the \texttt{CoolTrack} engine evaluates all localised surrogate models over a common sequence of internal temperatures, this initial formation entropy must be mapped back into temperature space to define the upper boundary of the evolutionary array. Given the inclusion of adaptive softplus parameterisations for entropy, we extracted the starting temperature ($T_0$) via a bounded scalar minimisation, bypassing the risk of numerical bracketing failures. In purely log-linear regimes, this process reduces to an instantaneous analytical inversion. 

The age ($t$) is then calculated by integrating the inverse cooling rate (residence time) over the physical entropy space:
\begin{equation}
    t = -\int_{S_0=S(T_0)}^{S(T_{\rm int})} \tau(S) \, \mathrm{d}S
.\end{equation}
While a purely log-linear formulation of the localised surrogates possesses an exact, closed-form analytical solution (provided as a first-order approximation in Appendix~\ref{sec:appendix_analytical_integral}), the dynamic inclusion of the softplus model to capture non-linear structural and cooling plateaus breaks this simple analytical form. Therefore, \texttt{CoolTrack} evaluates this integral numerically over the continuous analytical functions. 

Finally, we attached an uncertainty band to every surrogate. Although the bounded-softplus fits reproduce the grid well, their parameter covariances are near-degenerate, because the knee position and sharpness are only weakly identified, so sampling them produces unstable, runaway envelopes. We therefore characterised each surrogate by the proximity-weighted residual scatter of its own fit in log space,
\begin{equation}
    \sigma^2 = \frac{\sum_i w_i\, r_i^2}{\sum_i w_i},
\end{equation}
where $r_i$ is the residual of grid point $i$ and $w_i$ its Gaussian weight. Because $\sigma$ is defined in log space it is a constant fractional uncertainty, and the $1\sigma$ band on each quantity is $\hat{y}\,e^{\pm\sigma}$ (or $\hat{y}\pm\sigma$ for the photometry, fitted in $\log$ flux). This estimator tracks the fit quality, $\sigma \simeq \sqrt{1-R^2}\,\mathrm{std}(\ln y)$, so a quantity with a large dynamic range can retain a sizeable absolute band even at high $R^2$. For the age, the entropy and cooling-rate scatters are propagated coherently through $t = -\int \tau\,\mathrm{d}S$, offsetting each curve by its $\sigma$ before re-integrating; the radius scatter is evaluated independently, so the radius and age uncertainties remain orthogonal. The cooling-rate surrogate dominates the resulting age uncertainty, since the residence time spans the widest dynamic range.

\section{Results}

\subsection{Phase space validation}

As illustrated in Fig.~\ref{fig:surrogate_extraction}, the localised Gaussian weighting effectively isolates relevant physical regimes for a given target mass. This localised approach allows the mixed linear, softplus, and spline regressions to map the planet's complex thermodynamic and photometric states. The robustness of the softplus framework is particularly evident in the structural equation of state ($\ln R_{\text{eq}}$ vs $\ln T_{\rm int}$), where bounded fits yield weighted coefficients of determination ($R^2 \ge 0.86$). This formulation provides a stable, high-fidelity representation of the planetary contraction sequence across the entire mass domain, effectively capturing physical asymptotes while remaining resilient to numerical noise.

Beyond resolving deep structural limits, the surrogate framework provides critical mathematical stabilisation to the planet's thermal phase space. In traditional evolutionary modelling, age is derived by coupling an interior model to a pre-computed atmospheric grid and integrating the tabulated cooling derivative ($\mathrm{d}S/\mathrm{d}t$) via an ODE solver. However, these grids are constructed using 1D radiative-convective equilibrium codes that evaluate the atmosphere layer-by-layer. When a cooling planet crosses a critical condensation threshold, the sudden precipitation of thick cloud decks in specific atmospheric layers triggers strong opacity spikes and convective restructuring. While these codes eventually converge, they permanently imprint these sharp, non-linear gradients into the tabulated cooling derivatives. 

When a traditional ODE solver attempts to interpolate across these boundaries, it interprets these microscopic cloud-clearing events as numerical discontinuities, frequently resulting in step-size collapse and solver failure. The \texttt{CoolTrack} methodology inherently neutralises this stiffness. By projecting the discrete grid data into a logarithmic phase space and enforcing a strict top-of-atmosphere energy balance, the engine treats the planet as a bulk macroscopic system rather than being sensitive to the volatile internal layers. The adaptive softplus regression is then fitted directly over the macroscopic cooling sequence. This mathematical approach absorbs the underlying layer-by-layer atmospheric chaos into a single, continuously differentiable analytical curve. Consequently, the framework isolates and maps the true physical boundaries of the atmospheric cooling rate without ever exposing the integration engine to the raw numerical discontinuities of the discrete grid.

\subsection{Thermal, thermodynamic, and structural evolution}

\begin{figure*}[ht]
    \centering
    \includegraphics[width=\textwidth]{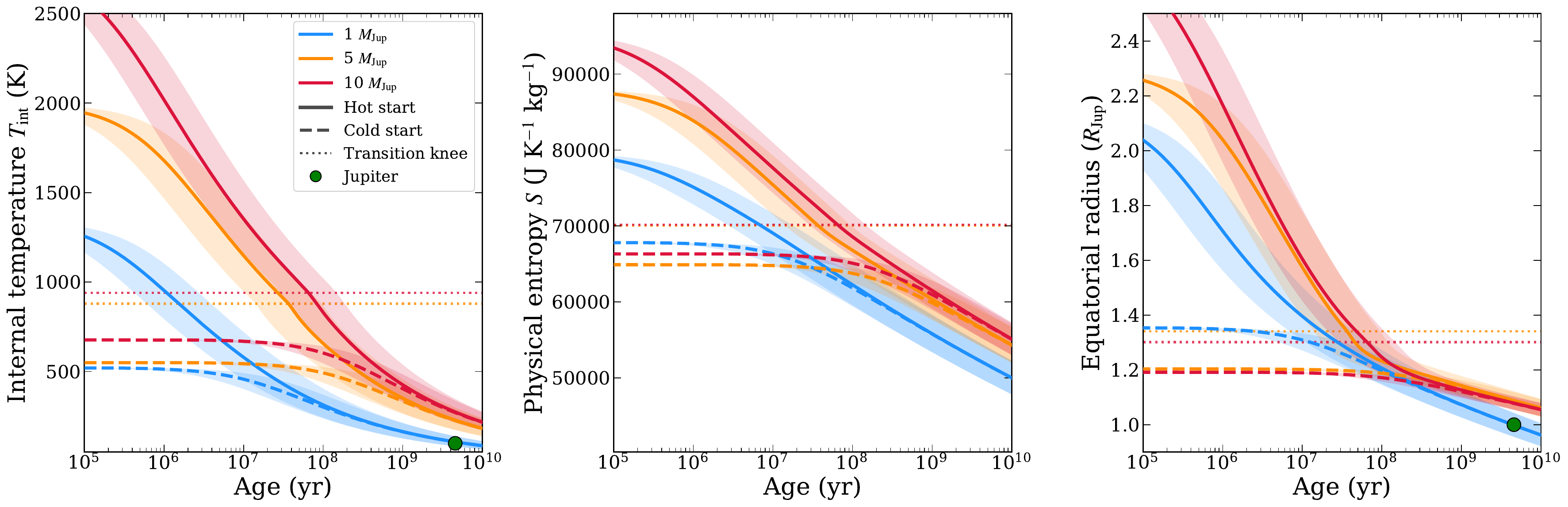}
    \caption{Continuous evolutionary tracks generated via numerical integration, comparing the thermal, thermodynamic, and structural evolution of 1.0, 5.0, and 10.0 $M_{\rm Jup}$ planets under a constant 100~K irradiation field. \textit{Left}: Thermal evolution mapping internal temperature ($T_{\rm int}$) against planetary age, highlighting the slope change during cloud clearing. \textit{Centre}: Thermodynamic evolution tracking physical entropy ($S$) against age. \textit{Right}: Structural evolution mapping equatorial radius against age. Solid and dashed lines denote hot- and cold-start limits, respectively. Shaded regions represent $1\sigma$ bands, taken for every surrogate as the proximity-weighted residual scatter of its local fit; the age band propagates the entropy and cooling-rate scatter through the integral, independently of the radius. The green circle indicates the present-day properties of Jupiter. The horizontal dotted lines correspond to the primary physical knees (identified via peak curvature in the softplus surrogate fits), denoting the critical temperatures, entropies, and radii at which the planets undergo their most dramatic structural or atmospheric transitions.}
\label{fig:evolution_mass_starts_3panel}
\end{figure*}

The numerical integration of the analytical surrogate models generates continuous and stable evolutionary tracks across the planetary mass domain, as illustrated in Fig.~\ref{fig:evolution_mass_starts_3panel}. By drawing the initial entropy bounds directly from our theoretical formation interpolators (see Sect.~\ref{Sec:fomula}), the semi-analytical engine rapidly produces the classic tuning fork trajectories for $1.0$, $5.0$, and $10.0~M_{\rm Jup}$ planets. These tracks effectively capture the divergence between high-entropy hot-start and low-entropy cold-start scenarios, demonstrating that giant planets retain a significant memory of their initial formation entropy for $10^7$--$10^8$~yr before converging onto a unified cooling sequence. The robustness of the framework is validated by its convergence on present-day Solar System benchmarks; for a $1~M_{\rm Jup}$ target, \texttt{CoolTrack} recovers the properties of Jupiter ($4.56 \times 10^9$~yr, $R \approx 1.0~R_{\rm Jup}$, $T_{\rm int} \approx 99$~K, the Voyager-era value; more recent analyses favour $\approx 107$~K; \citealt{li_less_2018}) without iterative ODE solvers.

A critical feature revealed by this three-panel mapping is the distinct change in cooling slope during atmospheric transitions. For a typical $10~M_{\rm Jup}$ object undergoing the L-to-T cloud-clearing sequence, the adaptive softplus architecture independently isolates distinct mathematical knees. The inflection occurs in $\tau$, which plateaus at $T_{\rm int} \approx 1009$~K. This represents the physical blanket removal phase, where the rapid rain-out of silicate and iron clouds shifts the surface boundary condition and forces the object to vent trapped heat. This manifests as a slope change of the thermal cooling curve (Fig.~\ref{fig:evolution_mass_starts_3panel}, left), similar to the statistical pile-up of transitional objects predicted by numerical models \citep{saumon_evolution_2008}.

However, as shown in the centre panel of Fig.~\ref{fig:evolution_mass_starts_3panel}, the deep interior does not instantaneously restructure itself. The bulk internal entropy ($S$) exhibits a structural-reaction knee a little later, at $T_{\rm int} \approx 953$~K. This $\sim 50$~K offset maps the physical thermodynamic lag: the planet must remain at an elevated temperature while the large heat deficit drains from the core. Only after tens of millions of years of venting does the internal entropy gradient fully adjust to its new, clear-sky cooling slope. By natively identifying this separation between surface emission and deep structural reaction, \texttt{CoolTrack} mathematically validates the temporal sequence of cloud-clearing without imposing rigid physical priors.

\subsection{Photometric light curves}
The semi-analytical surrogate framework translates the fundamental entropy evolution directly into observable photometric fluxes. By modelling the temporal photometric light curves with cubic B-splines, the engine captures the non-linear flux transitions that occur as the planet cools through critical cloud condensation sequences without triggering numerical solver instabilities. 

\begin{figure*}[ht]
    \centering
    \includegraphics[width=\textwidth]{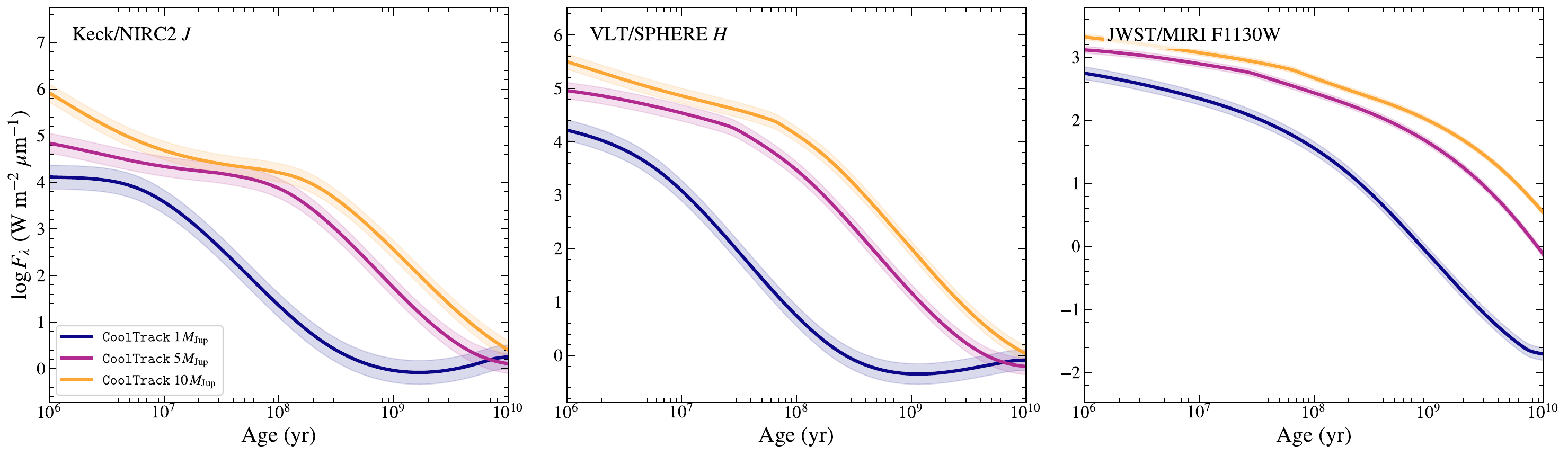}
    \caption{\texttt{CoolTrack} evolutionary light curves displaying unscaled surface fluxes ($\log$ W m$^{-2}$ $\mu$m$^{-1}$) for $1.0$, $5.0$, and $10.0~M_{\rm Jup}$ planets. The three panels illustrate emission across distinct observatory bands capturing diverse atmospheric physics: the steep L-to-T transition drop-off in Keck/NIRC2 J (\textit{left}), near-IR cloud-clearing in VLT/SPHERE IRDIS-H (\textit{centre}), and the smooth thermal decay settling onto the irradiation floor in JWST/MIRI F1130W (\textit{right}). The median tracks are generated via numerical integration of the cubic B-spline photometric regressions. The shaded regions represent $1\sigma$ bands taken as the proximity-weighted residual scatter of the spline fit in each band.}
    \label{fig:surface_flux_light_curves}
\end{figure*}

This continuous mapping is illustrated in Fig.~\ref{fig:surface_flux_light_curves}, which tracks the unscaled surface flux evolution across three distinct photometric bands: Keck/NIRC2 $J$ ($\lambda_{\text{eff}} \approx 1.25~\mu$m; Near-Infrared Camera 2), VLT/SPHERE $H$ ($\lambda_{\text{eff}} \approx 1.63~\mu$m), and JWST/MIRI F1130W ($\lambda_{\text{eff}} = 11.30~\mu$m; Mid-Infrared Instrument). These three bands were selected to showcase the engine's ability to model distinct physical regimes: the NIRC2 $J$ band traces the steep multi-dex drop-off of the L-to-T transition; the SPHERE $H$ band captures standard near-IR cloud-clearing; and the JWST MIRI F1130W band highlights the stable, deeply cooled thermal tail and its asymptotic settlement onto the background irradiation floor.

Because the B-spline is linear in its coefficients the fit is stable across the sharp cloud-clearing transitions, and we take the photometric uncertainty as the proximity-weighted residual scatter of the spline fit in each band, consistent with the structural and thermodynamic surrogates. This translates localised grid sparsity into the continuous noise envelopes visible in the shaded bounds of the light curves.

\subsection{Colour--magnitude evolution}
\begin{figure}[ht]
    \centering
    \includegraphics[width=\columnwidth]{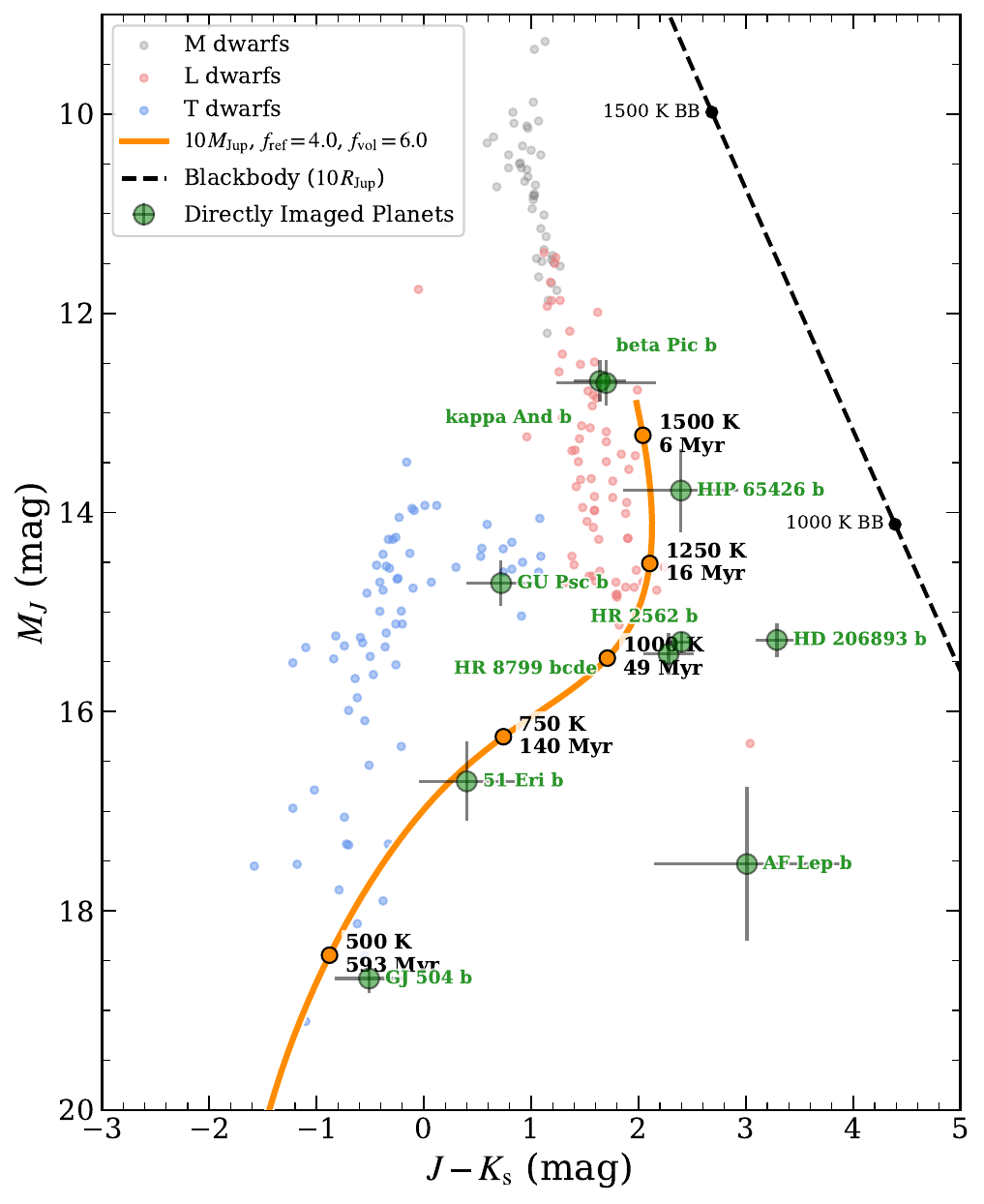}
    \caption{Evolutionary track of a $10~M_{\rm Jup}$ planet with moderate cloud sedimentation ($f_{\text{refractory}}=4.0$, $f_{\text{volatile}}=6.0$) in a $J-K_{\rm s}$ vs absolute $M_J$ colour--magnitude diagram. The solid orange line denotes the median \texttt{CoolTrack} surrogate prediction, with circular markers indicating specific internal temperatures and ages. The shaded region represents the $1\sigma$ residual-scatter band of the B-spline photometry. Background scatter points illustrate empirical M (grey), L (red), and T (blue) field dwarfs from the \texttt{species} toolkit \citep{stolker_miracles_2020}, while a theoretical $1~R_{\rm Jup}$ blackbody curve (dashed black line) is provided for reference. Directly imaged planetary-mass companions are overlaid as yellow stars: 51~Eri~b \citep{macintosh_discovery_2015, rajan_characterizing_2017}, GJ~504~b \citep{kuzuhara_direct_2013}, HR~8799~bcde \citep{zurlo_first_2016}, $\beta$~Pic~b \citep{currie_combined_2013}, $\kappa$~And~b \citep{carson_direct_2013}, GU~Psc~b \citep{naud_discovery_2014}, HR~2562~b \citep{konopacky_discovery_2016}, HIP~65426~b \citep{chauvin_discovery_2017}, HD~206893~b \citep{delorme_-depth_2017}, and AF~Lep~b \citep{gratton_implications_2024}. The \texttt{CoolTrack} track and the blackbody are computed in the Keck/NIRC2 $J$ and $K_{\rm s}$ bands; the field-dwarf sequence and the directly imaged companions combine near-IR photometry from heterogeneous systems (MKO $K$ / 2MASS $K_{\rm s}$) whose $\lesssim 0.1$~mag system differences are negligible compared with the plotted uncertainties and the intrinsic scatter of the sequences.}
    \label{fig:cmd_evolution}
\end{figure}

To validate the photometric surrogates against empirical observations, we evaluate the \texttt{CoolTrack} engine in colour--magnitude space. Figure~\ref{fig:cmd_evolution} presents the evolutionary trajectory of a $10~M_{\rm Jup}$ planet in an absolute $M_J$ versus $J-K_{\rm s}$ colour--magnitude diagram. The background distributions represent field M, L, and T dwarfs drawn from the Database of Ultracool Parallaxes \citep{dupuy_hawaii_2012}, alongside explicitly observed planetary-mass wide companions.

As the planet cools from its initial hot-start entropy, the B-spline photometric regressions smoothly navigate the complex atmospheric phase space. At high temperatures ($T_{\rm int} > 1200$~K), the thick refractory clouds maintain a red $J-K_{\rm s}$ colour, closely tracking the empirical L-dwarf sequence. As the planet cools through the critical L-to-T transition, the volatile cloud decks condense and sink beneath the photosphere. The surrogate captures this sharp, non-linear atmospheric clearing event, driving the evolutionary track bluewards into the T-dwarf regime. Crucially, the localised spline regression models the high-temperature $J$-band hook caused by the shifting peak of the Planck function without introducing artificial extrapolation artefacts. The sharp transition being modelled at lower J magnitude than the observed field brown dwarfs can be related to the study of planetary mass objects with lower 1-bar gravities than brown dwarfs, as shown by \citet{charnay_self-consistent_2018}.

\section{Discussion}

\subsection{Current limitations and missing physics}
\label{Sec:limitations}
While the current \texttt{CoolTrack} framework bypasses ODE solver instabilities across broad evolutionary tracks, the accuracy of the surrogate models is inherently tied to the completeness of the underlying atmospheric and structural grids. At present, the extracted surrogates natively accommodate the transition from ideal gas contraction to electron degeneracy as a fully parameterised possibility within the framework. However, there remain specific physical phenomena that are not yet explicitly parameterised.

For instance, mature, deeply cooled giant planets (such as Saturn) undergo hydrogen--helium (H/He) phase separation in their deep interiors. The rain-out of helium droplets releases latent heat and gravitational energy, significantly suppressing the cooling rate and delaying the planet's evolution \citep{howard_evolution_2024}. 
A more fundamental simplification is shared with the literature models against which we benchmark. The Sonora \citep{marley_sonora_2021}, ATMO~2020 \citep{phillips_new_2020}, \citet{baraffe_evolutionary_2003} and \citet{burrows_nongray_1997} tracks, and the helium-rain models of \citet{howard_evolution_2024} all assume a fully convective, adiabatic interior of uniform composition. The interiors revealed by Juno and Cassini are not so simple: both Jupiter and Saturn appear to host dilute, extended fuzzy cores with stably stratified composition gradients across which heat is transported non-adiabatically \citep{knierim_convective_2024, tejada_arevalo_jupiter_2024}. Modelling such interiors requires evolution codes that integrate the coupled structure and composition explicitly, such as \texttt{APPLE} \citep{sur_apple_2024} and \texttt{ORCHARD} \citep{arevalo_orchard_2026}, and simultaneous fits to Jupiter and Saturn already show the imprint of fuzzy cores on the inferred thermal history \citep{sur_simultaneous_2025, sur_next-generation_2026}. Because \texttt{CoolTrack} inherits the thermodynamics of its underlying grid, an adiabatic, fully convective \texttt{HADES} grid cannot reproduce these non-adiabatic pathways; the surrogate could in principle ingest a grid built with stratified interiors, but the present results apply to the adiabatic, fully convective regime. Similarly, planets subjected to extreme stellar irradiation environments can experience severe photodissociation of atmospheric molecules, creating strong temperature inversions that alter the top-of-atmosphere energy balance and subsequent cooling pathways \citep{parmentier_thermal_2018, lothringer_extremely_2018}.

A major advantage of our semi-analytical approach is its structural adaptability. Because we do not rely on a rigid step-by-step numerical integrator, missing physics that introduce abrupt changes to the cooling derivative ($\mathrm{d}S/\mathrm{d}t$) do not break the engine. For instance, the entropy surrogate already stacks a second softplus segment to capture the low-temperature volatile inflection below $\sim 300$~K, and the same mechanism extends to further cooling plateaus by adding segments to the formulation. Similarly, to account for H--He phase separation, the current equations can be updated by introducing an additional knee at the critical temperature threshold for helium immiscibility, allowing the surrogate to analytically capture the sudden flattening of the cooling curve without introducing numerical grid artefacts. Similarly, the effects of severe atmospheric photodissociation, which frequently trigger upper-atmospheric temperature inversions in extreme stellar environments, can be parameterised by introducing localised sigmoid terms. Because these inversions typically manifest across specific irradiation thresholds, a sigmoid function could dynamically adjust the top-of-atmosphere boundary conditions along the $T_{\rm irr}$ axis, capturing the suppressed cooling pathway without introducing numerical discontinuities.

\subsection{Usage cases and Bayesian inversions}
The primary operational advantage of the \texttt{CoolTrack} engine is its high computational efficiency, which effectively removes the forward-modelling bottleneck inherent in planetary evolution studies. This makes the surrogate framework uniquely suited for direct integration within atmospheric retrieval pipelines and Bayesian MCMC inversion codes, where evaluating $10^5$ to $10^6$ models is routinely required. A recent, practical demonstration of this capability is the characterisation of the circumbinary planet HD 143811 (AB)b \citep{squicciarini_gpisphere_2025}. In that study, the \texttt{HADES} grid and surrogate evolutionary framework were embedded directly into the MCMC atmospheric retrieval. Rather than relying on a fixed, assumed age to determine the planet's mass and radius, the system's age was treated as a dynamic parameter constrained by soft boundaries. The retrieval utilised a soft upper limit based on the host star's isochronal age ($\sim 21$~Myr), consistent with the system's kinematic membership of the young Scorpius--Centaurus association \citep{squicciarini_gpisphere_2025}, and a soft lower limit based on the median survival time of protoplanetary discs ($\sim 5$~Myr; \citealt{haisch_jr_disk_2001}). By dynamically marginalising over these age and structural uncertainties within the surrogate framework, the resulting posteriors are expected to represent the actual parameter uncertainties more faithfully than traditional fixed-parameter cooling models.

Furthermore, because the semi-analytical engine requires the initial formation entropy ($S_0$) as a boundary condition, it allows observers to directly explore specific formation mechanisms within the retrieval itself. By altering the parameterised starting entropy, the MCMC can shift between testing hot-start (gravitational collapse), cold-start (core accretion), or intermediate warm-start formation scenarios, explicitly linking the observed atmospheric photometry to the planet's primordial formation history. Importantly, when independent dynamical mass constraints are available (e.g. via astrometry or radial velocity), this framework is uniquely positioned to perform comparisons between formation scenarios, directly testing which primordial entropy states most consistently reconcile the planet's gravitational mass with its observed photometric evolution.

\subsection{Synergy with machine learning and physics-informed neural networks}

While universal function approximators, such as neural networks composed of weighted sums of sigmoids, are capable of fitting complex multidimensional data, the physically driven approach of \texttt{CoolTrack} offers superior interpretability and control. By utilising analytical functions like the softplus to explicitly map physical inflections (e.g. the effect of clouds), we preserve a direct link between the surrogate parameters and the underlying planetary physics. However, the mathematical stability and speed of this framework make it an ideal foundation for future machine learning integration. The continuous, physically consistent evolutionary tracks generated by \texttt{CoolTrack} could serve as high-fidelity training data for neural networks designed to infer direct or transit spectra. This hybrid approach would essentially create a physics-informed neural network (PINN), where the machine learning model is trained on data that natively adhere to the thermodynamic and structural constraints of the semi-analytical engine, ensuring that spectral predictions remain grounded in realistic planetary evolution.

\section{Conclusions}
We have demonstrated that the temporal evolution of giant planets can be reproduced using local semi-analytical surrogate models. By replacing the traditional ODE solver with log-log Gaussian-weighted linear regressions, softplus structural asymptotes, and cubic B-spline photometric regressions, we achieve mathematically stable contraction curves and light curves.

Furthermore, because it replaces the computationally expensive ODE solver with vectorised numerical integration over continuous analytical functions, \texttt{CoolTrack} is computationally efficient. Benchmarking tests demonstrate that once the localised analytical parameters are defined, a full 500-node evolutionary track, spanning thermodynamics, cooling rates, structural radii, and complex multi-band photometry, is evaluated and integrated in $\sim 0.15$ milliseconds on a standard desktop CPU (e.g. a single Apple M2 / Intel i7 core). While the initial extraction of these parameters via distance-weighted regression scales with the size of the underlying atmospheric grid, this bottleneck is bypassed in production. By pre-mapping the surrogate coefficients across the target parameter space prior to runtime, the engine maintains millisecond performance regardless of grid density. This efficiency removes the forward-modelling bottleneck inherent in planetary evolution studies, making the surrogate framework uniquely suited for use within atmospheric retrieval pipelines and Bayesian MCMC inversion codes, where $10^5$ to $10^6$ sequential model evaluations are routinely required.

\section*{Code and data availability}
\texttt{CoolTrack} is implemented in Python (tested with CPython~3.12; requires Python~$\ge 3.8$) and built on NumPy, SciPy (\texttt{optimize}, \texttt{interpolate}, \texttt{integrate}, \texttt{spatial}), scikit-learn and pandas, with pyarrow for grid input/output and Matplotlib for visualisation. It is distributed as an installable package together with a pre-compiled coefficient database, so evolutionary tracks can be generated without loading the underlying \texttt{HADES} grids, and with tutorial notebooks reproducing the figures in this paper. Track generation runs on a single core of a standard desktop CPU and requires no GPU. The code and coefficient database are available at \url{https://github.com/ChristianSWilkinson/cooltrack}.

\begin{acknowledgements}
We thank the referee, Ankan Sur, for a constructive report that improved the manuscript. This project has received funding from the European Research Council (ERC) under the European Union’s Horizon 2020 research and innovation programme (COBREX; grant agreement n° 885593). This work was granted access to the HPC resources of MesoPSL financed by the Region Ile de France and the project Equip@Meso (reference ANR-10-EQPX-29-01) of the programme Investissements d’Avenir supervised by the Agence Nationale pour la Recherche. This work was supported by CNES, focused on AIRS on Ariel, and by the Programme National de Planétologie (PNP) of CNRS/INSU, cofunded by CNES. The authors acknowledge the use of Google's Gemini (version 3.1 Pro, \url{https://gemini.google.com}) for assistance with code development, brainstorming, language editing, and grammatical polishing. Special thanks to the open-source community for the SciPy \citep{virtanen_scipy_2020} and NumPy libraries. 
\end{acknowledgements}

\bibliographystyle{aa} 
\bibliography{library} 

@article{phillips_new_2020,
    author = {Phillips, Mark W. and Tremblin, Pascal and Baraffe, Isabelle and Chabrier, Gilles and Allard, Nicole F. and Spiegelman, Fernand and Goyal, Jayesh M. and Drummond, Ben and Hebrard, Eric},
    title = {A new set of atmosphere and evolution models for cool {T}-{Y} brown dwarfs and giant exoplanets},
    journal = {A\&A},
    year = {2020},
    volume = {637},
    pages = {A38},
    doi = {10.1051/0004-6361/201937381},
}

@article{marley_sonora_2021,
    author = {Marley, Mark S. and Saumon, Didier and Visscher, Channon and Lupu, Roxana and Freedman, Richard and Morley, Caroline and Fortney, Jonathan J. and Seay, Christopher and Smith, Adam J. R. W. and Teal, D. J. and Wang, Ruoyan},
    title = {The {Sonora} {Brown} {Dwarf} {Atmosphere} and {Evolution} {Models} {I}. {Model} {Description} and {Application} to {Cloudless} {Atmospheres} in {Rainout} {Chemical} {Equilibrium}},
    journal = {ApJ},
    year = {2021},
    volume = {920},
    number = {2},
    pages = {85},
    doi = {10.3847/1538-4357/ac141d},
}

@article{marley_luminosity_2007,
    author = {Marley, M. S. and Fortney, J. J. and Hubickyj, O. and Bodenheimer, P. and Lissauer, J. J.},
    title = {On the {Luminosity} of {Young} {Jupiters}},
    journal = {ApJ},
    year = {2007},
    volume = {655},
    number = {1},
    pages = {541--549},
    doi = {10.1086/509759},
}

@article{baudino_interpreting_2015,
    author = {Baudino, J.-L. and B\'{e}zard, B. and Boccaletti, A. and Bonnefoy, M. and Lagrange, A.-M. and Galicher, R.},
    title = {Interpreting the photometry and spectroscopy of directly imaged planets: a new atmospheric model applied to \textit{$\beta$} {Pictoris} b and {SPHERE} observations},
    journal = {A\&A},
    year = {2015},
    volume = {582},
    pages = {A83},
    doi = {10.1051/0004-6361/201526332},
}

@article{chabrier_new_2021,
    author = {Chabrier, Gilles and Debras, Florian},
    title = {A new equation of state for dense hydrogen-helium mixtures {II}: taking into account hydrogen-helium interactions},
    journal = {ApJ},
    year = {2021},
    volume = {917},
    number = {1},
    pages = {4},
    doi = {10.3847/1538-4357/abfc48},
}

@article{mordasini_characterization_2017,
    author = {Mordasini, C. and Marleau, G.-D. and Molli\`{e}re, P.},
    title = {Characterization of exoplanets from their formation: {III}. {The} statistics of planetary luminosities},
    journal = {A\&A},
    year = {2017},
    volume = {608},
    pages = {A72},
    doi = {10.1051/0004-6361/201630077},
}

@article{baraffe_evolutionary_2003,
    author = {Baraffe, I. and Chabrier, G. and Barman, T. and Allard, F. and Hauschildt, P. H.},
    title = {Evolutionary models for cool brown dwarfs and extrasolar giant planets. {The} case of {HD} 20945},
    journal = {A\&A},
    year = {2003},
    volume = {402},
    number = {2},
    pages = {701--712},
    doi = {10.1051/0004-6361:20030252},
}

@article{lagrange_probable_2009,
    author = {Lagrange, A.-M. and Gratadour, D. and Chauvin, G. and Fusco, T. and Ehrenreich, D. and Mouillet, D. and Rousset, G. and Rouan, D. and Allard, F. and Gendron, E. and Charton, J. and Mugnier, L. and Rabou, P. and Montri, J. and Lacombe, F.},
    title = {A probable giant planet imaged in the {Beta} {Pictoris} disk},
    journal = {A\&A},
    year = {2009},
    volume = {493},
    number = {2},
    pages = {L21--L25},
    doi = {10.1051/0004-6361:200811325},
}

@article{marois_direct_2008,
    author = {Marois, C. and Macintosh, B. and Barman, T. and Zuckerman, B. and Song, I. and Patience, J. and Lafreniere, D. and Doyon, R.},
    title = {Direct {Imaging} of {Multiple} {Planets} {Orbiting} the {Star} {HR} 8799},
    journal = {Science},
    year = {2008},
    volume = {322},
    number = {5906},
    pages = {1348--1352},
    doi = {10.1126/science.1166585},
}

@article{macintosh_discovery_2015,
    author = {Macintosh, B. and Graham, J. R. and Barman, T. and De Rosa, R. J. and Konopacky, Q. and Marley, M. S. and Marois, C. and Nielsen, E. L. and Pueyo, L. and Rajan, A. and Rameau, J. and Saumon, D. and Wang, J. J. and Patience, J. and Ammons, M. and Arriaga, P. and Artigau, E. and Beckwith, S. and Brewster, J. and Bruzzone, S. and Bulger, J. and Burningham, B. and Burrows, A. S. and Chen, C. and Chiang, E. and Chilcote, J. K. and Dawson, R. I. and Dong, R. and Doyon, R. and Draper, Z. H. and Duch\^{e}ne, G. and Esposito, T. M. and Fabrycky, D. and Fitzgerald, M. P. and Follette, K. B. and Fortney, J. J. and Gerard, B. and Goodsell, S. and Greenbaum, A. Z. and Hibon, P. and Hinkley, S. and Cotten, T. H. and Hung, L.-W. and Ingraham, P. and Johnson-Groh, M. and Kalas, P. and Lafreniere, D. and Larkin, J. E. and Lee, J. and Line, M. and Long, D. and Maire, J. and Marchis, F. and Matthews, B. C. and Max, C. E. and Metchev, S. and Millar-Blanchaer, M. A. and Mittal, T. and Morley, C. V. and Morzinski, K. M. and Murray-Clay, R. and Oppenheimer, R. and Palmer, D. W. and Patel, R. and Perrin, M. D. and Poyneer, L. A. and Rafikov, R. R. and Rantakyr\"{o}, F. T. and Rice, E. L. and Rojo, P. and Rudy, A. R. and Ruffio, J.-B. and Ruiz, M. T. and Sadakuni, N. and Saddlemyer, L. and Salama, M. and Savransky, D. and Schneider, A. C. and Sivaramakrishnan, A. and Song, I. and Soummer, R. and Thomas, S. and Vasisht, G. and Wallace, J. K. and Ward-Duong, K. and Wiktorowicz, S. J. and Wolff, S. G. and Zuckerman, B.},
    title = {Discovery and spectroscopy of the young {Jovian} planet 51 {Eri} b with the {Gemini} {Planet} {Imager}},
    journal = {Science},
    year = {2015},
    volume = {350},
    number = {6256},
    pages = {64--67},
    doi = {10.1126/science.aac5891},
}

@article{charnay_self-consistent_2018,
    author = {Charnay, B. and B\'{e}zard, B. and Baudino, J.-L. and Bonnefoy, M. and Boccaletti, A. and Galicher, R.},
    title = {A {Self}-consistent {Cloud} {Model} for {Brown} {Dwarfs} and {Young} {Giant} {Exoplanets}: {Comparison} with {Photometric} and {Spectroscopic} {Observations}},
    journal = {ApJ},
    year = {2018},
    volume = {854},
    number = {2},
    pages = {172},
    doi = {10.3847/1538-4357/aaac7d},
}

@article{rajan_characterizing_2017,
    author = {Rajan, Abhijith and Rameau, Julien and De Rosa, Robert J. and Marley, Mark S. and Graham, James R. and Macintosh, Bruce and Marois, Christian and Morley, Caroline and Patience, Jennifer and Pueyo, Laurent and Saumon, Didier and Ward-Duong, Kimberly and Ammons, S. Mark and Arriaga, Pauline and Bailey, Vanessa P. and Barman, Travis and Bulger, Joanna and Burrows, Adam S. and Chilcote, Jeffrey and Cotten, Tara and Czekala, Ian and Doyon, Rene and Duch\^{e}ne, Gaspard and Esposito, Thomas M. and Fitzgerald, Michael P. and Follette, Katherine B. and Fortney, Jonathan J. and Goodsell, Stephen J. and Greenbaum, Alexandra Z. and Hibon, Pascale and Hung, Li-Wei and Ingraham, Patrick and Johnson-Groh, Mara and Kalas, Paul and Konopacky, Quinn and Lafreni\`{e}re, David and Larkin, James E. and Maire, J\'{e}r\^{o}me and Marchis, Franck and Metchev, Stanimir and Millar-Blanchaer, Maxwell A. and Morzinski, Katie M. and Nielsen, Eric L. and Oppenheimer, Rebecca and Palmer, David and Patel, Rahul I. and Perrin, Marshall and Poyneer, Lisa and Rantakyr\"{o}, Fredrik T. and Ruffio, Jean-Baptiste and Savransky, Dmitry and Schneider, Adam C. and Sivaramakrishnan, Anand and Song, Inseok and Soummer, R\'{e}mi and Thomas, Sandrine and Vasisht, Gautam and Wallace, J. Kent and Wang, Jason J. and Wiktorowicz, Sloane and Wolff, Schuyler},
    title = {Characterizing 51 {Eri} b from 1-5 \${\textbackslash}mu\$m: a partly-cloudy exoplanet},
    journal = {AJ},
    year = {2017},
    volume = {154},
    number = {1},
    pages = {10},
    doi = {10.3847/1538-3881/aa74db},
}

@article{saumon_evolution_2008,
    author = {Saumon, D. and Marley, M. S.},
    title = {The {Evolution} of {L} and {T} {Dwarfs} in {Color}-{Magnitude} {Diagrams}},
    journal = {ApJ},
    year = {2008},
    volume = {689},
    number = {2},
    pages = {1327--1344},
    doi = {10.1086/592734},
}

@article{ackerman_precipitating_2001,
    author = {Ackerman, Andrew S. and Marley, Mark S.},
    title = {Precipitating {Condensation} {Clouds} in {Substellar} {Atmospheres}},
    journal = {ApJ},
    year = {2001},
    volume = {556},
    number = {2},
    pages = {872--884},
    doi = {10.1086/321540},
}

@article{haisch_jr_disk_2001,
    author = {Haisch, Jr., Karl E. and Lada, Elizabeth A. and Lada, Charles J.},
    title = {Disk {Frequencies} and {Lifetimes} in {Young} {Clusters}},
    journal = {ApJ},
    year = {2001},
    volume = {553},
    number = {2},
    pages = {L153--L156},
    doi = {10.1086/320685},
}

@article{dupuy_hawaii_2012,
    author = {Dupuy, Trent J. and Liu, Michael C.},
    title = {{THE} {HAWAII} {INFRARED} {PARALLAX} {PROGRAM}. {I}. {ULTRACOOL} {BINARIES} {AND} {THE} {L}/{T} {TRANSITION} $^{\textrm{,}}$},
    journal = {ApJS},
    year = {2012},
    volume = {201},
    number = {2},
    pages = {19},
    doi = {10.1088/0067-0049/201/2/19},
}

@article{carson_direct_2013,
    author = {Carson, J. and Thalmann, C. and Janson, M. and Kozakis, T. and Bonnefoy, M. and Biller, B. and Schlieder, J. and Currie, T. and McElwain, M. and Goto, M. and Henning, T. and Brandner, W. and Feldt, M. and Kandori, R. and Kuzuhara, M. and Stevens, L. and Wong, P. and Gainey, K. and Fukagawa, M. and Kuwada, Y. and Brandt, T. and Kwon, J. and Abe, L. and Egner, S. and Grady, C. and Guyon, O. and Hashimoto, J. and Hayano, Y. and Hayashi, M. and Hayashi, S. and Hodapp, K. and Ishii, M. and Iye, M. and Knapp, G. and Kudo, T. and Kusakabe, N. and Matsuo, T. and Miyama, S. and Morino, J. and Moro-Martin, A. and Nishimura, T. and Pyo, T. and Serabyn, E. and Suto, H. and Suzuki, R. and Takami, M. and Takato, N. and Terada, H. and Tomono, D. and Turner, E. and Watanabe, M. and Wisniewski, J. and Yamada, T. and Takami, H. and Usuda, T. and Tamura, M.},
    title = {{DIRECT} {IMAGING} {DISCOVERY} {OF} {A} ``{SUPER}-{JUPITER}'' {AROUND} {THE} {LATE} {B}-{TYPE} {STAR} \textbf{ \textit{$\kappa$} } {And}},
    journal = {ApJ},
    year = {2013},
    volume = {763},
    number = {2},
    pages = {L32},
    doi = {10.1088/2041-8205/763/2/L32},
}

@article{currie_combined_2013,
    author = {Currie, Thayne and Burrows, Adam and Madhusudhan, Nikku and Fukagawa, Misato and Girard, Julien H. and Dawson, Rebekah and Murray-Clay, Ruth and Kenyon, Scott and Kuchner, Marc and Matsumura, Soko and Jayawardhana, Ray and Chambers, John and Bromley, Ben},
    title = {A {COMBINED} {VERY} {LARGE} {TELESCOPE} {AND} {GEMINI} {STUDY} {OF} {THE} {ATMOSPHERE} {OF} {THE} {DIRECTLY} {IMAGED} {PLANET}, $\beta$ {PICTORIS} b},
    journal = {ApJ},
    year = {2013},
    volume = {776},
    number = {1},
    pages = {15},
    doi = {10.1088/0004-637X/776/1/15},
}

@article{kuzuhara_direct_2013,
    author = {Kuzuhara, M. and Tamura, M. and Kudo, T. and Janson, M. and Kandori, R. and Brandt, T. D. and Thalmann, C. and Spiegel, D. and Biller, B. and Carson, J. and Hori, Y. and Suzuki, R. and Burrows, A. and Henning, T. and Turner, E. L. and McElwain, M. W. and Moro-Mart\'{\i}n, A. and Suenaga, T. and Takahashi, Y. H. and Kwon, J. and Lucas, P. and Abe, L. and Brandner, W. and Egner, S. and Feldt, M. and Fujiwara, H. and Goto, M. and Grady, C. A. and Guyon, O. and Hashimoto, J. and Hayano, Y. and Hayashi, M. and Hayashi, S. S. and Hodapp, K. W. and Ishii, M. and Iye, M. and Knapp, G. R. and Matsuo, T. and Mayama, S. and Miyama, S. and Morino, J.-I. and Nishikawa, J. and Nishimura, T. and Kotani, T. and Kusakabe, N. and Pyo, T.-S. and Serabyn, E. and Suto, H. and Takami, M. and Takato, N. and Terada, H. and Tomono, D. and Watanabe, M. and Wisniewski, J. P. and Yamada, T. and Takami, H. and Usuda, T.},
    title = {{DIRECT} {IMAGING} {OF} {A} {COLD} {JOVIAN} {EXOPLANET} {IN} {ORBIT} {AROUND} {THE} {SUN}-{LIKE} {STAR} {GJ} 504},
    journal = {ApJ},
    year = {2013},
    volume = {774},
    number = {1},
    pages = {11},
    doi = {10.1088/0004-637X/774/1/11},
}

@article{naud_discovery_2014,
    author = {Naud, Marie-Eve and Artigau, \'{E}tienne and Malo, Lison and Albert, Lo\"{\i}c and Doyon, Ren\'{e} and Lafreni\`{e}re, David and Gagn\'{e}, Jonathan and Saumon, Didier and Morley, Caroline V. and Allard, France and Homeier, Derek and Beichman, Charles A. and Gelino, Christopher R. and Boucher, Anne},
    title = {{DISCOVERY} {OF} {A} {WIDE} {PLANETARY}-{MASS} {COMPANION} {TO} {THE} {YOUNG} {M3} {STAR} {GU} {PSC}},
    journal = {ApJ},
    year = {2014},
    volume = {787},
    number = {1},
    pages = {5},
    doi = {10.1088/0004-637X/787/1/5},
}

@article{konopacky_discovery_2016,
    author = {Konopacky, Quinn M. and Rameau, Julien and Duch\^{e}ne, Gaspard and Filippazzo, Joseph C. and Godfrey, Paige A. Giorla and Marois, Christian and Nielsen, Eric L. and Pueyo, Laurent and Rafikov, Roman R. and Rice, Emily L. and Wang, Jason J. and Ammons, S. Mark and Bailey, Vanessa P. and Barman, Travis S. and Bulger, Joanna and Bruzzone, Sebastian and Chilcote, Jeffrey K. and Cotten, Tara and Dawson, Rebekah I. and Rosa, Robert J. De and Doyon, Ren\'{e} and Esposito, Thomas M. and Fitzgerald, Michael P. and Follette, Katherine B. and Goodsell, Stephen and Graham, James R. and Greenbaum, Alexandra Z. and Hibon, Pascale and Hung, Li-Wei and Ingraham, Patrick and Kalas, Paul and Lafreni\`{e}re, David and Larkin, James E. and Macintosh, Bruce A. and Maire, J\'{e}r\^{o}me and Marchis, Franck and Marley, Mark S. and Matthews, Brenda C. and Metchev, Stanimir and Millar-Blanchaer, Maxwell A. and Oppenheimer, Rebecca and Palmer, David W. and Patience, Jenny and Perrin, Marshall D. and Poyneer, Lisa A. and Rajan, Abhijith and Rantakyr\"{o}, Fredrik T. and Savransky, Dmitry and Schneider, Adam C. and Sivaramakrishnan, Anand and Song, Inseok and Soummer, Remi and Thomas, Sandrine and Wallace, J. Kent and Ward-Duong, Kimberly and Wiktorowicz, Sloane J. and Wolff, Schuyler G.},
    title = {{DISCOVERY} {OF} {A} {SUBSTELLAR} {COMPANION} {TO} {THE} {NEARBY} {DEBRIS} {DISK} {HOST} {HR} 2562},
    journal = {ApJL},
    year = {2016},
    volume = {829},
    number = {1},
    pages = {L4},
    doi = {10.3847/2041-8205/829/1/L4},
}

@article{zurlo_first_2016,
    author = {Zurlo, A. and Vigan, A. and Galicher, R. and Maire, A.-L. and Mesa, D. and Gratton, R. and Chauvin, G. and Kasper, M. and Moutou, C. and Bonnefoy, M. and Desidera, S. and Abe, L. and Apai, D. and Baruffolo, A. and Baudoz, P. and Baudrand, J. and Beuzit, J.-L. and Blancard, P. and Boccaletti, A. and Cantalloube, F. and Carle, M. and Charton, J. and Claudi, R. U. and Costille, A. and de Caprio, V. and Dohlen, K. and Dominik, C. and Fantinel, D. and Feautrier, P. and Feldt, M. and Fusco, T. and Gascone, E. and Gigan, P. and Girard, J. H. and Gissler, D. and Gluck, L. and Gry, C. and Henning, T. and Hugot, E. and Janson, M. and Jacquet, M. and Lagrange, A.-M. and Langlois, M. and Llored, M. and Made, F. and Magnard, Y. and Martinez, P. and Maurel, D. and Mawet, D. and Meyer, M. R. and Milli, J. and Moeller-Nilsson, O. and Mouillet, D. and Orign\'{e}, A. and Pavlov, A. and Petit, C. and Puget, P. and Quanz, S. P. and Rabou, P. and Ramos, J. and Roux, A. and Salasnich, B. and Salter, G. and Sauvage, J.-F. and Schmid, H. M. and Soenke, C. and Stadler, E. and Suarez, M. and Turatto, M. and Udry, S. and Vakili, F. and Wahhaj, Z. and Wildi, F.},
    title = {First light of the {VLT} planet finder {SPHERE}. {III}. {New} spectrophotometry and astrometry of the {HR8799} exoplanetary system},
    journal = {A\&A},
    year = {2016},
    volume = {587},
    pages = {A57},
    doi = {10.1051/0004-6361/201526835},
}

@article{delorme_-depth_2017,
    author = {Delorme, P. and Schmidt, T. and Bonnefoy, M. and Desidera, S. and Ginski, C. and Charnay, B. and Lazzoni, C. and Christiaens, V. and Messina, S. and D'Orazi, V. and Milli, J. and Schlieder, J. E. and Gratton, R. and Rodet, L. and Lagrange, A.-M. and Absil, O. and Vigan, A. and Galicher, R. and Hagelberg, J. and Bonavita, M. and Lavie, B. and Zurlo, A. and Olofsson, J. and Boccaletti, A. and Cantalloube, F. and Mouillet, D. and Chauvin, G. and Hambsch, F.-J. and Langlois, M. and Udry, S. and Henning, T. and Beuzit, J.-L. and Mordasini, C. and Lucas, P. and Marocco, F. and Biller, B. and Carson, J. and Cheetham, A. and Covino, E. and De Caprio, V. and Delboulbe, A. and Feldt, M. and Girard, J. and Hubin, N. and Maire, A.-L. and Pavlov, A. and Petit, C. and Rouan, D. and Roelfsema, R. and Wildi, F.},
    title = {In-depth study of moderately young but extremely red, very dusty substellar companion {HD} {206893B}},
    journal = {A\&A},
    year = {2017},
    volume = {608},
    pages = {A79},
    doi = {10.1051/0004-6361/201731145},
}

@article{chauvin_discovery_2017,
    author = {Chauvin, G. and Desidera, S. and Lagrange, A.-M. and Vigan, A. and Gratton, R. and Langlois, M. and Bonnefoy, M. and Beuzit, J.-L. and Feldt, M. and Mouillet, D. and Meyer, M. and Cheetham, A. and Biller, B. and Boccaletti, A. and D'Orazi, V. and Galicher, R. and Hagelberg, J. and Maire, A.-L. and Mesa, D. and Olofsson, J. and Samland, M. and Schmidt, T. O. B. and Sissa, E. and Bonavita, M. and Charnay, B. and Cudel, M. and Daemgen, S. and Delorme, P. and Janin-Potiron, P. and Janson, M. and Keppler, M. and Le Coroller, H. and Ligi, R. and Marleau, G. D. and Messina, S. and Molli\`{e}re, P. and Mordasini, C. and M\"{u}ller, A. and Peretti, S. and Perrot, C. and Rodet, L. and Rouan, D. and Zurlo, A. and Dominik, C. and Henning, T. and Menard, F. and Schmid, H.-M. and Turatto, M. and Udry, S. and Vakili, F. and Abe, L. and Antichi, J. and Baruffolo, A. and Baudoz, P. and Baudrand, J. and Blanchard, P. and Bazzon, A. and Buey, T. and Carbillet, M. and Carle, M. and Charton, J. and Cascone, E. and Claudi, R. and Costille, A. and Deboulbe, A. and De Caprio, V. and Dohlen, K. and Fantinel, D. and Feautrier, P. and Fusco, T. and Gigan, P. and Giro, E. and Gisler, D. and Gluck, L. and Hubin, N. and Hugot, E. and Jaquet, M. and Kasper, M. and Madec, F. and Magnard, Y. and Martinez, P. and Maurel, D. and Le Mignant, D. and M\"{o}ller-Nilsson, O. and Llored, M. and Moulin, T. and Orign\'{e}, A. and Pavlov, A. and Perret, D. and Petit, C. and Pragt, J. and Puget, P. and Rabou, P. and Ramos, J. and Rigal, R. and Rochat, S. and Roelfsema, R. and Rousset, G. and Roux, A. and Salasnich, B. and Sauvage, J.-F. and Sevin, A. and Soenke, C. and Stadler, E. and Suarez, M. and Weber, L. and Wildi, F. and Antoniucci, S. and Augereau, J.-C. and Baudino, J.-L. and Brandner, W. and Engler, N. and Girard, J. and Gry, C. and Kral, Q. and Kopytova, T. and Lagadec, E. and Milli, J. and Moutou, C. and Schlieder, J. and Szul\'{a}gyi, J. and Thalmann, C. and Wahhaj, Z.},
    title = {Discovery of a warm, dusty giant planet around {HIP} 65426},
    journal = {A\&A},
    year = {2017},
    volume = {605},
    pages = {L9},
    doi = {10.1051/0004-6361/201731152},
}

@article{burrows_nongray_1997,
    author = {Burrows, A. and Marley, M. and Hubbard, W. B. and Lunine, J. I. and Guillot, T. and Saumon, D. and Freedman, R. and Sudarsky, D. and Sharp, C.},
    title = {A {Nongray} {Theory} of {Extrasolar} {Giant} {Planets} and {Brown} {Dwarfs}},
    journal = {ApJ},
    year = {1997},
    volume = {491},
    number = {2},
    pages = {856--875},
    doi = {10.1086/305002},
}

@article{gratton_implications_2024,
    author = {Gratton, R. and Bonavita, M. and Mesa, D. and Zurlo, A. and Marino, S. and Desidera, S. and D'Orazi, V. and Rigliaco, E. and Squicciarini, V. and Nogueira, P. H.},
    title = {Implications of the discovery of {AF} {Lep} b: {The} mass-luminosity relation for planets in the \textit{$\beta$} {Pic} {Moving} {Group} and the {L}--{T} transition for young companions and free-floating planets},
    journal = {A\&A},
    year = {2024},
    volume = {684},
    pages = {A69},
    doi = {10.1051/0004-6361/202348012},
}

@article{stolker_miracles_2020,
    author = {Stolker, T. and Quanz, S. P. and Todorov, K. O. and K\"{u}hn, J. and Molli\`{e}re, P. and Meyer, M. R. and Currie, T. and Daemgen, S. and Lavie, B.},
    title = {{MIRACLES}: atmospheric characterization of directly imaged planets and substellar companions at 4--5 \textit{$\mu$} m: {I}. {Photometric} analysis of \textit{$\beta$} {Pic} b, {HIP} 65426 b, {PZ} {Tel} {B}, and {HD} 206893 {B}},
    journal = {A\&A},
    year = {2020},
    volume = {635},
    pages = {A182},
    doi = {10.1051/0004-6361/201937159},
}

@article{chabrier_theory_2000,
    author = {Chabrier, G. and Baraffe, I.},
    title = {Theory of {Low}-{Mass} {Stars} and {Substellar} {Objects}},
    journal = {ARA\&A},
    year = {2000},
    volume = {38},
    number = {1},
    pages = {337--377},
    doi = {10.1146/annurev.astro.38.1.337},
}

@article{al-refaie_taurex_2021,
    author = {Al-Refaie, Ahmed F. and Changeat, Quentin and Waldmann, Ingo P. and Tinetti, Giovanna},
    title = {{TauREx} {III}: {A} fast, dynamic and extendable framework for retrievals},
    journal = {ApJ},
    year = {2021},
    volume = {917},
    number = {1},
    pages = {37},
    doi = {10.3847/1538-4357/ac0252},
}

@article{tejada_arevalo_jupiter_2024,
    author = {Tejada Arevalo, R. and Sur, A. and Su, Y. and Burrows, A.},
    title = {Jupiter {Evolutionary} {Models} {Incorporating} {Stably} {Stratified} {Regions}},
    journal = {ApJ},
    year = {2025},
    volume = {979},
    pages = {243},
    doi = {10.3847/1538-4357/ada030},
}

@article{wilkinson_breaking_2024,
    author = {Wilkinson, C. and Charnay, B. and Mazevet, S. and Lagrange, A.-M. and Chomez, A. and Squicciarini, V. and Panek, E. and Mazoyer, J.},
    title = {Breaking degeneracies in exoplanetary parameters through self-consistent atmosphere--interior modelling},
    journal = {A\&A},
    year = {2024},
    volume = {692},
    pages = {A113},
    doi = {10.1051/0004-6361/202348945},
}

@article{lagrange_evidence_2025,
    author = {Lagrange, A.-M. and Wilkinson, C. and M\^{a}lin, M. and Boccaletti, A. and Perrot, C. and others},
    title = {Evidence for a sub-{J}ovian planet in the young {TWA} 7 disk},
    journal = {Nature},
    year = {2025},
    volume = {642},
    pages = {905},
    doi = {10.1038/s41586-025-09150-4},
}

@article{himes_accurate_2022,
    author = {Himes, Michael D. and Harrington, Joseph and Cobb, Adam D. and Baydin, Atilim Gunes and Soboczenski, Frank and O'Beirne, Molly D. and Zorzan, Simone and Wright, David C. and Scheffer, Zacchaeus and Domagal-Goldman, Shawn D. and Arney, Giada N.},
    title = {Accurate {Machine} {Learning} {Atmospheric} {Retrieval} via a {Neural} {Network} {Surrogate} {Model} for {Radiative} {Transfer}},
    journal = {PSJ},
    year = {2022},
    volume = {3},
    number = {4},
    pages = {91},
    doi = {10.3847/PSJ/abe3fd},
}

@article{madhusudhan_temperature_2009,
    author = {Madhusudhan, N. and Seager, S.},
    title = {A {Temperature} and {Abundance} {Retrieval} {Method} for {Exoplanet} {Atmospheres}},
    journal = {ApJ},
    year = {2009},
    volume = {707},
    number = {1},
    pages = {24--39},
    doi = {10.1088/0004-637X/707/1/24},
}

@article{squicciarini_gpisphere_2025,
    author = {Squicciarini, V. and Mazoyer, J. and Wilkinson, C. and Lagrange, A.-M. and Delorme, P. and Radcliffe, A. and Flasseur, O. and Kiefer, F. and Alecian, E.},
    title = {{GPI}+{SPHERE} detection of a 6.1 $M_{\rm Jup}$ circumbinary planet around {HD} 143811},
    journal = {A\&A},
    year = {2025},
    volume = {702},
    pages = {L10},
    doi = {10.1051/0004-6361/202557104},
}

@article{virtanen_scipy_2020,
    author = {Virtanen, Pauli and Gommers, Ralf and Oliphant, Travis E. and Haberland, Matt and Reddy, Tyler and Cournapeau, David and Burovski, Evgeni and Peterson, Pearu and Weckesser, Warren and Bright, Jonathan and Van Der Walt, St\'{e}fan J. and Brett, Matthew and Wilson, Joshua and Millman, K. Jarrod and Mayorov, Nikolay and Nelson, Andrew R. J. and Jones, Eric and Kern, Robert and Larson, Eric and Carey, C J and Polat, \.{I}lhan and Feng, Yu and Moore, Eric W. and VanderPlas, Jake and Laxalde, Denis and Perktold, Josef and Cimrman, Robert and Henriksen, Ian and Quintero, E. A. and Harris, Charles R. and Archibald, Anne M. and Ribeiro, Ant\^{o}nio H. and Pedregosa, Fabian and Van Mulbregt, Paul and {SciPy 1.0 Contributors} and Vijaykumar, Aditya and Bardelli, Alessandro Pietro and Rothberg, Alex and Hilboll, Andreas and Kloeckner, Andreas and Scopatz, Anthony and Lee, Antony and Rokem, Ariel and Woods, C. Nathan and Fulton, Chad and Masson, Charles and H\"{a}ggstr\"{o}m, Christian and Fitzgerald, Clark and Nicholson, David A. and Hagen, David R. and Pasechnik, Dmitrii V. and Olivetti, Emanuele and Martin, Eric and Wieser, Eric and Silva, Fabrice and Lenders, Felix and Wilhelm, Florian and Young, G. and Price, Gavin A. and Ingold, Gert-Ludwig and Allen, Gregory E. and Lee, Gregory R. and Audren, Herv\'{e} and Probst, Irvin and Dietrich, J\"{o}rg P. and Silterra, Jacob and Webber, James T and Slavi\v{c}, Janko and Nothman, Joel and Buchner, Johannes and Kulick, Johannes and Sch\"{o}nberger, Johannes L. and De Miranda Cardoso, Jos\'{e} Vin\'{\i}cius and Reimer, Joscha and Harrington, Joseph and Rodr\'{\i}guez, Juan Luis Cano and Nunez-Iglesias, Juan and Kuczynski, Justin and Tritz, Kevin and Thoma, Martin and Newville, Matthew and K\"{u}mmerer, Matthias and Bolingbroke, Maximilian and Tartre, Michael and Pak, Mikhail and Smith, Nathaniel J. and Nowaczyk, Nikolai and Shebanov, Nikolay and Pavlyk, Oleksandr and Brodtkorb, Per A. and Lee, Perry and McGibbon, Robert T. and Feldbauer, Roman and Lewis, Sam and Tygier, Sam and Sievert, Scott and Vigna, Sebastiano and Peterson, Stefan and More, Surhud and Pudlik, Tadeusz and Oshima, Takuya and Pingel, Thomas J. and Robitaille, Thomas P. and Spura, Thomas and Jones, Thouis R. and Cera, Tim and Leslie, Tim and Zito, Tiziano and Krauss, Tom and Upadhyay, Utkarsh and Halchenko, Yaroslav O. and V\'{a}zquez-Baeza, Yoshiki},
    title = {{SciPy} 1.0: fundamental algorithms for scientific computing in {Python}},
    journal = {Nat. Methods},
    year = {2020},
    volume = {17},
    number = {3},
    pages = {261--272},
    doi = {10.1038/s41592-019-0686-2},
}

@article{knierim_convective_2024,
    author = {Knierim, Henrik and Helled, Ravit},
    title = {Convective {Mixing} in {Gas} {Giant} {Planets} with {Primordial} {Composition} {Gradients}},
    journal = {ApJ},
    year = {2024},
    volume = {977},
    number = {2},
    pages = {227},
    doi = {10.3847/1538-4357/ad8dd0},
}

@article{sur_apple_2024,
    author = {Sur, Ankan and Su, Yubo and Tejada Arevalo, Roberto and Chen, Yi-Xian and Burrows, Adam},
    title = {{APPLE}: {An} {Evolution} {Code} for {Modeling} {Giant} {Planets}},
    journal = {ApJ},
    year = {2024},
    volume = {971},
    number = {1},
    pages = {104},
    doi = {10.3847/1538-4357/ad57c3},
}

@article{cleveland_locally_1988,
    author = {Cleveland, William S. and Devlin, Susan J.},
    title = {Locally {Weighted} {Regression}: {An} {Approach} to {Regression} {Analysis} by {Local} {Fitting}},
    journal = {J. Am. Stat. Assoc.},
    year = {1988},
    volume = {83},
    number = {403},
    pages = {596--610},
    publisher = {Taylor \& Francis},
    doi = {10.1080/01621459.1988.10478639},
}

@article{de_wringer_surrogate-accelerated_2026,
    author = {De Wringer, Tijn and Dorn, Caroline and Garvin, Emily O. and Marelli, Stefano},
    title = {Surrogate-accelerated {Bayesian} {Inversion} for {Exoplanet} {Interior} {Characterization}},
    journal = {ApJ},
    year = {2026},
    volume = {997},
    number = {2},
    pages = {321},
    doi = {10.3847/1538-4357/ae2ec4},
}

@article{stadter_benchmarking_2021,
    author = {St\"{a}dter, Philipp and Sch\"{a}lte, Yannik and Schmiester, Leonard and Hasenauer, Jan and Stapor, Paul L.},
    title = {Benchmarking of numerical integration methods for {ODE} models of biological systems},
    journal = {Sci. Rep.},
    year = {2021},
    volume = {11},
    number = {1},
    pages = {2696},
    doi = {10.1038/s41598-021-82196-2},
}

@article{creswell_understanding_2023,
    author = {Creswell, R. and Shepherd, K. M. and Lambert, B. and Mirams, G. R. and Lei, C. L. and Tavener, S. and Robinson, M. and Gavaghan, D. J.},
    title = {Understanding the impact of numerical solvers on inference for differential equation models},
    journal = {J. R. Soc. Interface},
    year = {2024},
    volume = {21},
    pages = {20230369},
    doi = {10.1098/rsif.2023.0369},
}

@article{burgasser_spectra_2002,
    author = {Burgasser, Adam J. and Kirkpatrick, J. Davy and Brown, Michael E. and Reid, I. Neill and Burrows, Adam and Liebert, James and Matthews, Keith and Gizis, John E. and Dahn, Conard C. and Monet, David G. and Cutri, Roc M. and Skrutskie, Michael F.},
    title = {The {Spectra} of {T} {Dwarfs} {I}: {Near}-{Infrared} {Data} and {Spectral} {Classification}},
    journal = {ApJ},
    year = {2002},
    volume = {564},
    number = {1},
    pages = {421--451},
    doi = {10.1086/324033},
}

@article{howard_evolution_2024,
    author = {Howard, Saburo and M\"{u}ller, Simon and Helled, Ravit},
    title = {Evolution of {Jupiter} and {Saturn} with helium rain},
    journal = {A\&A},
    year = {2024},
    volume = {689},
    pages = {A15},
    doi = {10.1051/0004-6361/202450629},
}

@article{parmentier_thermal_2018,
    author = {Parmentier, Vivien and Line, Mike R. and Bean, Jacob L. and Mansfield, Megan and Kreidberg, Laura and Lupu, Roxana and Visscher, Channon and D\'{e}sert, Jean-Michel and Fortney, Jonathan J. and Deleuil, Magalie and Arcangeli, Jacob and Showman, Adam P. and Marley, Mark S.},
    title = {From thermal dissociation to condensation in the atmospheres of ultra hot {Jupiters}: {WASP}-121b in context},
    journal = {A\&A},
    year = {2018},
    volume = {617},
    pages = {A110},
    doi = {10.1051/0004-6361/201833059},
}

@article{lothringer_extremely_2018,
    author = {Lothringer, Joshua D. and Barman, Travis and Koskinen, Tommi},
    title = {Extremely {Irradiated} {Hot} {Jupiters}: {Non}-oxide {Inversions}, {H}$^{\textrm{-}}$ {Opacity}, and {Thermal} {Dissociation} of {Molecules}},
    journal = {ApJ},
    year = {2018},
    volume = {866},
    number = {1},
    pages = {27},
    doi = {10.3847/1538-4357/aadd9e},
}

@article{lodders_alkali_1999,
    author = {Lodders, Katharina},
    title = {Alkali {Element} {Chemistry} in {Cool} {Dwarf} {Atmospheres}},
    journal = {ApJ},
    year = {1999},
    volume = {519},
    number = {2},
    pages = {793--801},
    doi = {10.1086/307387},
}

@incollection{mason_chemistry_2006,
    author = {Lodders, K. and Fegley, B.},
    editor = {Mason, John W.},
    title = {Chemistry of {Low} {Mass} {Substellar} {Objects}},
    booktitle = {Astrophysics {Update} 2},
    year = {2006},
    pages = {1--28},
    publisher = {Springer Berlin Heidelberg},
    address = {Berlin, Heidelberg},
    doi = {10.1007/3-540-30313-8_1},
}

@article{visscher_atmospheric_2010,
    author = {Visscher, Channon and Lodders, Katharina and Fegley, Bruce},
    title = {{ATMOSPHERIC} {CHEMISTRY} {IN} {GIANT} {PLANETS}, {BROWN} {DWARFS}, {AND} {LOW}-{MASS} {DWARF} {STARS}. {III}. {IRON}, {MAGNESIUM}, {AND} {SILICON}},
    journal = {ApJ},
    year = {2010},
    volume = {716},
    number = {2},
    pages = {1060--1075},
    doi = {10.1088/0004-637X/716/2/1060},
}

@article{morley_thermal_2015,
    author = {Morley, Caroline V. and Fortney, Jonathan J. and Marley, Mark S. and Zahnle, Kevin and Line, Michael and Kempton, Eliza and Lewis, Nikole and Cahoy, Kerri},
    title = {{THERMAL} {EMISSION} {AND} {REFLECTED} {LIGHT} {SPECTRA} {OF} {SUPER} {EARTHS} {WITH} {FLAT} {TRANSMISSION} {SPECTRA}},
    journal = {ApJ},
    year = {2015},
    volume = {815},
    number = {2},
    pages = {110},
    doi = {10.1088/0004-637X/815/2/110},
}

@article{mckay_comparison_1979,
    author = {Mckay, M. and Beckman, Richard and Conover, William},
    title = {A {Comparison} of {Three} {Methods} for {Selecting} {Vales} of {Input} {Variables} in the {Analysis} of {Output} {From} a {Computer} {Code}},
    journal = {Technometrics},
    year = {1979},
    volume = {21},
    pages = {239--245},
    doi = {10.1080/00401706.1979.10489755},
}

@article{macdonald_poseidon_2023,
    author = {MacDonald, Ryan J.},
    title = {{POSEIDON}: {A} {Multidimensional} {Atmospheric} {Retrieval} {Code} for {Exoplanet} {Spectra}},
    journal = {J. Open Source Softw.},
    year = {2023},
    volume = {8},
    number = {81},
    pages = {4873},
    doi = {10.21105/joss.04873},
}

@article{molliere_petitradtrans_2019,
    author = {Molli\`{e}re, P. and Wardenier, J. P. and Van Boekel, R. and Henning, Th. and Molaverdikhani, K. and Snellen, I. A. G.},
    title = {{petitRADTRANS}: {A} {Python} radiative transfer package for exoplanet characterization and retrieval},
    journal = {A\&A},
    year = {2019},
    volume = {627},
    pages = {A67},
    doi = {10.1051/0004-6361/201935470},
}

@article{ardevol_martinez_floppity_2024,
    author = {Ard\'{e}vol Mart\'{\i}nez, F. and Min, M. and Huppenkothen, D. and Kamp, I. and Palmer, P. I.},
    title = {{FlopPITy}: {Enabling} self-consistent exoplanet atmospheric retrievals with machine learning},
    journal = {A\&A},
    year = {2024},
    volume = {681},
    pages = {L14},
    doi = {10.1051/0004-6361/202348367},
}

@incollection{chabrier_giant_2014,
    author = {Chabrier, G. and Johansen, A. and Janson, M. and Rafikov, R.},
    title = {Giant planet and brown dwarf formation},
    booktitle = {Protostars and {Planets} {VI}},
    year = {2014},
    publisher = {University of Arizona Press},
    doi = {10.2458/azu_uapress_9780816531240-ch027},
}

@article{bowler_imaging_2016,
    author = {Bowler, Brendan P.},
    title = {Imaging {Extrasolar} {Giant} {Planets}},
    journal = {PASP},
    year = {2016},
    volume = {128},
    number = {968},
    pages = {102001},
    doi = {10.1088/1538-3873/128/968/102001},
}

@article{sur_next-generation_2026,
    author = {Sur, Ankan and {Tejada Arevalo}, Roberto and Burrows, Adam and Chen, Yi-Xian},
    title = {Next-generation {Improvements} in {Giant}-exoplanet {Evolutionary} and {Structural} {Models}},
    journal = {ApJ},
    year = {2026},
    volume = {998},
    number = {2},
    pages = {305},
    doi = {10.3847/1538-4357/ae3a85},
}

@misc{arevalo_orchard_2026,
    author = {{Tejada Arevalo}, Roberto and Burrows, Adam and Sur, Ankan and Su, Yubo},
    title = {{ORCHARD}: {A} {General} {Planetary} {Evolution} {Code}},
    howpublished = {arXiv:2604.24845},
    year = {2026},
    publisher = {arXiv},
    doi = {10.48550/ARXIV.2604.24845},
    eprint = {2604.24845},
    archiveprefix = {arXiv},
}

@article{li_less_2018,
    author = {Li, Liming and Jiang, X. and West, R. A. and Gierasch, P. J. and Perez-Hoyos, S. and Sanchez-Lavega, A. and Fletcher, L. N. and Fortney, J. J. and Knowles, B. and Porco, C. C. and Baines, K. H. and Fry, P. M. and Mallama, A. and Achterberg, R. K. and Simon, A. A. and Nixon, C. A. and Orton, G. S. and Dyudina, U. A. and Ewald, S. P. and Schmude, R. W.},
    title = {Less absorbed solar energy and more internal heat for {Jupiter}},
    journal = {Nat. Commun.},
    year = {2018},
    volume = {9},
    number = {1},
    pages = {3709},
    doi = {10.1038/s41467-018-06107-2},
}

@article{sur_simultaneous_2025,
    author = {Sur, Ankan and Tejada Arevalo, Roberto and Su, Yubo and Burrows, Adam},
    title = {Simultaneous {Evolutionary} {Fits} for {Jupiter} and {Saturn} {Incorporating} {Fuzzy} {Cores}},
    journal = {ApJL},
    year = {2025},
    volume = {980},
    number = {1},
    pages = {L5},
    doi = {10.3847/2041-8213/adad62},
}

@article{curtiss_integration_1952,
    author = {Curtiss, C. F. and Hirschfelder, J. O.},
    title = {Integration of {Stiff} {Equations}},
    journal = {PNAS},
    year = {1952},
    volume = {38},
    number = {3},
    pages = {235--243},
    doi = {10.1073/pnas.38.3.235},
}

@article{caballero_review_2018,
    author = {Caballero, Jos\'{e} A.},
    title = {A {Review} on {Substellar} {Objects} below the {Deuterium} {Burning} {Mass} {Limit}: {Planets}, {Brown} {Dwarfs} or {What}?},
    journal = {Geosciences},
    year = {2018},
    volume = {8},
    number = {10},
    pages = {362},
    doi = {10.3390/geosciences8100362},
}

@article{chen_jupiter_2023,
    author = {Chen, Yi-Xian and Burrows, Adam and Sur, Ankan and {Tejada Arevalo}, Roberto},
    title = {Jupiter {Atmospheric} {Models} and {Outer} {Boundary} {Conditions} for {Giant} {Planet} {Evolutionary} {Calculations}},
    journal = {ApJ},
    year = {2023},
    volume = {957},
    number = {1},
    pages = {36},
    doi = {10.3847/1538-4357/acf456},
}

@article{spiegel_deuterium-burning_2011,
    author = {Spiegel, David S. and Burrows, Adam and Milsom, John A.},
    title = {{THE} {DEUTERIUM}-{BURNING} {MASS} {LIMIT} {FOR} {BROWN} {DWARFS} {AND} {GIANT} {PLANETS}},
    journal = {ApJ},
    year = {2011},
    volume = {727},
    number = {1},
    pages = {57},
    doi = {10.1088/0004-637X/727/1/57},
}

@article{kirkpatrick_new_2005,
    author = {Kirkpatrick, J. Davy},
    title = {New {Spectral} {Types} {L} and {T}},
    journal = {ARA\&A},
    year = {2005},
    volume = {43},
    pages = {195--245},
    publisher = {Annual Reviews},
    doi = {10.1146/annurev.astro.42.053102.134017},
}

@article{luhman_formation_2012,
    author = {Luhman, Kevin L.},
    title = {The {Formation} and {Early} {Evolution} of {Low}-mass {Stars} and {Brown} {Dwarfs}},
    journal = {ARA\&A},
    year = {2012},
    volume = {50},
    number = {1},
    pages = {65--106},
    doi = {10.1146/annurev-astro-081811-125528},
}

@article{osorio_discovery_2000,
    author = {{Zapatero Osorio}, M. R. and B\'{e}jar, V. J. S. and Mart\'{\i}n, E. L. and Rebolo, R. and {Barrado y Navascu\'{e}s}, D. and Bailer-Jones, C. A. L. and Mundt, R.},
    title = {Discovery of {Young}, {Isolated} {Planetary} {Mass} {Objects} in the $\sigma$ {Orionis} {Star} {Cluster}},
    journal = {Science},
    year = {2000},
    volume = {290},
    number = {5489},
    pages = {103--107},
    doi = {10.1126/science.290.5489.103},
}

\begin{appendix}
\nolinenumbers
\section{Analytical derivation of the age integral}
\label{sec:appendix_analytical_integral}

The derivation below provides a closed-form analytical solution for planetary age as a first-order baseline approximation, applicable strictly when the thermodynamics and cooling rates operate purely in the log-linear regime. Note that for full evolutionary tracks, the inclusion of adaptive softplus transitions necessitates the numerical integration utilised by the \texttt{CoolTrack} engine.

Within the purely log-linear bounds of the surrogate model, the physical entropy ($S$) and inverse cooling rate ($\tau$) are represented by the formulations: \begin{align}
    S(T_{\rm int}) &= \exp(\beta_S) \cdot T_{\rm int}^{\alpha_S} \label{eq:S_exp} \\
    \tau(T_{\rm int}) &= \exp(\beta_\tau) \cdot T_{\rm int}^{\alpha_\tau} \label{eq:tau_exp}
\end{align}
where the dependences on the static planetary parameter vector $\boldsymbol{\theta}$ are omitted for brevity. The temporal evolution (age $t$) is defined by the integration of the inverse cooling rate over the entropy space:
\begin{equation}
    t = -\int_{S_0}^{S} \tau \, dS
.\end{equation}
To evaluate this analytically, we differentiated Eq.~\ref{eq:S_exp} with respect to $T_{\rm int}$:
\begin{equation}
    dS = \alpha_S \exp(\beta_S) \cdot T_{\rm int}^{\alpha_S - 1} \, dT_{\rm int}
.\end{equation}
Substituting this and Eq.~\ref{eq:tau_exp} into the age integral, and changing the integration limits to the corresponding initial ($T_0$) and current ($T_{\rm int}$) internal temperatures, yields
\begin{equation}
    t = -\int_{T_0}^{T_{\rm int}} \left[ \exp(\beta_\tau) \cdot T^{\alpha_\tau} \right] \left[ \alpha_S \exp(\beta_S) \cdot T^{\alpha_S - 1} \right] \, dT
.\end{equation}
Consolidating constants and exponents,
\begin{equation}
    t = - \alpha_S \exp(\beta_S + \beta_\tau) \int_{T_0}^{T_{\rm int}} T^{(\alpha_S + \alpha_\tau - 1)} \, dT
.\end{equation}
Integrating with respect to $T$ provides the exact analytical formulation for the age of the planet strictly as a function of internal temperature and the localised regression coefficients:
\begin{equation}
    t(T_{\rm int}) = \frac{\alpha_S \exp(\beta_S + \beta_\tau)}{\alpha_S + \alpha_\tau} \left( T_0^{(\alpha_S + \alpha_\tau)} - T_{\rm int}^{(\alpha_S + \alpha_\tau)} \right)
.\end{equation}
Because $T_{\rm int} < T_0$ for a cooling substellar object, the parenthetical term is negative, which cancels the leading negative sign implicit in the cooling derivation, resulting in a strictly positive age. Furthermore, a mathematical singularity where $\alpha_S + \alpha_\tau = 0$ is physically impossible in the substellar regime. While $\alpha_S$ is positive (entropy decreases as the object cools) and $\alpha_\tau$ is negative (cooling slows down over time), the magnitude of the cooling deceleration vastly outweighs the rate of entropy loss ($|\alpha_\tau| \gg \alpha_S$), ensuring the denominator remains non-zero.
\end{appendix}
\end{document}